\documentclass[prl, twocolumn, showpacs, nofootinbib, amsmath,amssymb, floatfix, eqsecnum]{revtex4-1}
\usepackage{amsmath}
\usepackage{braket}
\usepackage{amssymb}
\usepackage{amsthm}
\usepackage{algpseudocode}
\usepackage{algorithm}
\usepackage{amsfonts}
\usepackage{comment}
\usepackage[normalem]{ulem}
\usepackage{graphicx}
\usepackage{color,framed}
\usepackage{hyperref}
\usepackage{times}
\usepackage{enumerate}
\usepackage{lipsum}
\usepackage{slashed}
\usepackage{url}
\usepackage{bbm}
\usepackage{chngcntr}
\counterwithout{equation}{section}
\usepackage{tikz,pgfplots}
\usepackage{pgfplotstable}
\usepackage{siunitx}
\usepackage{comment}
\usepackage{graphicx}

\makeatletter
\newcommand{\setlabel}[1]{\edef\@currentlabel{#1}\label}
\makeatother

\usepgfplotslibrary{fillbetween}

\hypersetup{
    colorlinks=true, 
    linktoc=all,     
    linkcolor=blue,  
}

\def \beq {\begin{equation}}
\def \eeq {\end{equation}}
\def \beqa {\begin{eqnarray}}
\def \eeqa {\end{eqnarray}}
\def \bseq {\begin{subequations}}
\def \eseq {\end{subequations}}

\newcommand{\D}[1]{\text{d}#1}

\pgfplotsset{compat=1.18}
\begin{document}

 \title{Practicality of quantum adiabatic algorithm for chemistry applications}
\author{Etienne Granet, Khaldoon Ghanem and Henrik Dreyer}
\affiliation{Quantinuum, Leopoldstrasse 180, 80804 Munich, Germany}

\begin{abstract}
Despite its simplicity and strong theoretical guarantees, adiabatic state preparation has received considerably less interest than variational approaches for the preparation of low-energy electronic structure states. Two major reasons for this are the large number of gates required for Trotterising time-dependent electronic structure Hamiltonians, as well as discretisation errors heating the state. We show that a recently proposed randomized algorithm \cite{granet2023continuous}, which implements exact adiabatic evolution without heating and with far fewer gates than Trotterisation, can overcome this problem. We develop three methods for measuring the energy of the prepared state in an efficient and noise-resilient manner, yielding chemically accurate results on a 4-qubit molecule in the presence of realistic gate noise, without the need for error mitigation. These findings suggest that adiabatic approaches to state preparation could play a key role in quantum chemistry simulations both in the era of noisy as well as error-corrected quantum computers.
\end{abstract}

\maketitle

\section{I. Introduction}

Representing low-energy states of the electronic structure Hamiltonian is an essential step in quantum chemistry. Initial proposals for the use of quantum computers for this task have used Quantum Phase Estimation to project high-overlap Hartree-Fock wavefunctions into the ground state with high probability \cite{aspuru2005simulated}. The high cost of the coherent time-evolution required, as well as the existence of strongly multireference ground states (i.e.~molecules where the Hartree-Fock state has low overlap with the ground state) has triggered an explosion of interest in how to efficiently prepare a low-energy electronic structure state on a digital quantum computer.

Heuristic approaches, most notably the variational quantum eigensolver (VQE), have been devised and carried out on noisy devices~\cite{peruzzo2014variational,mcclean2016theory,tilly2022variational}. While inspired by adiabatic evolution~\cite{farhi2000quantum}, these methods abandon the theoretical guarantees of the adiabatic approach in favor of parameterized circuits which need to be optimized. For the classical computer, this replaces the quantum simulation problem with the usually hard problem of non-linear optimization in a generically hostile cost landscape~\cite{mcclean2018barren,cerezo2021cost,wang2021noise,anschuetz2022quantum,dreyer2021quantum}.

In contrast, adiabatic approaches to the state preparation problem have received considerably less attention, despite more theoretical guarantees and fewer hidden cost than VQE~\cite{aspuru2005simulated,veis2014adiabatic,lee2023evaluating,babbush2014adiabatic,kremenetski2021simulation,du2010nmr,bauer2016hybrid,lee2023evaluating,hayasaka2023general}. A major reasons for this is that the adiabatic evolution is usually assumed to be implemented via Trotterisation, in part because more sophisticated methods suffer from extra overheads~\cite{childs2012hamiltonian,berry2015simulating,berry2014exponential,low2017optimal,low2019hamiltonian,low2018hamiltonian,low2016methodology,kikuchi2023realization}, especially when simulating time-dependent Hamiltonians~\cite{poulin2011quantum,berry2014exponential,berry2015simulating,kieferova2019simulating,low2018hamiltonian,berry2020time}. Trotterising the electronic structure Hamiltonian, which contains $O(L^4)$ terms where $L$ is the number of orbitals, then requires extremely large numbers of gates. What's more, discretising a continuous adiabatic path into piecewise constant Trotter steps necessarily leads to discretisation errors (Trotter errors), which effectively heat the quantum state. This is particularly problematic in quantum chemistry applications, where extreme precision is required to have predictive power on chemical reaction rates, called chemical accuracy.

\begin{figure}[h]
    \centering
    \includegraphics[width=0.45\textwidth]{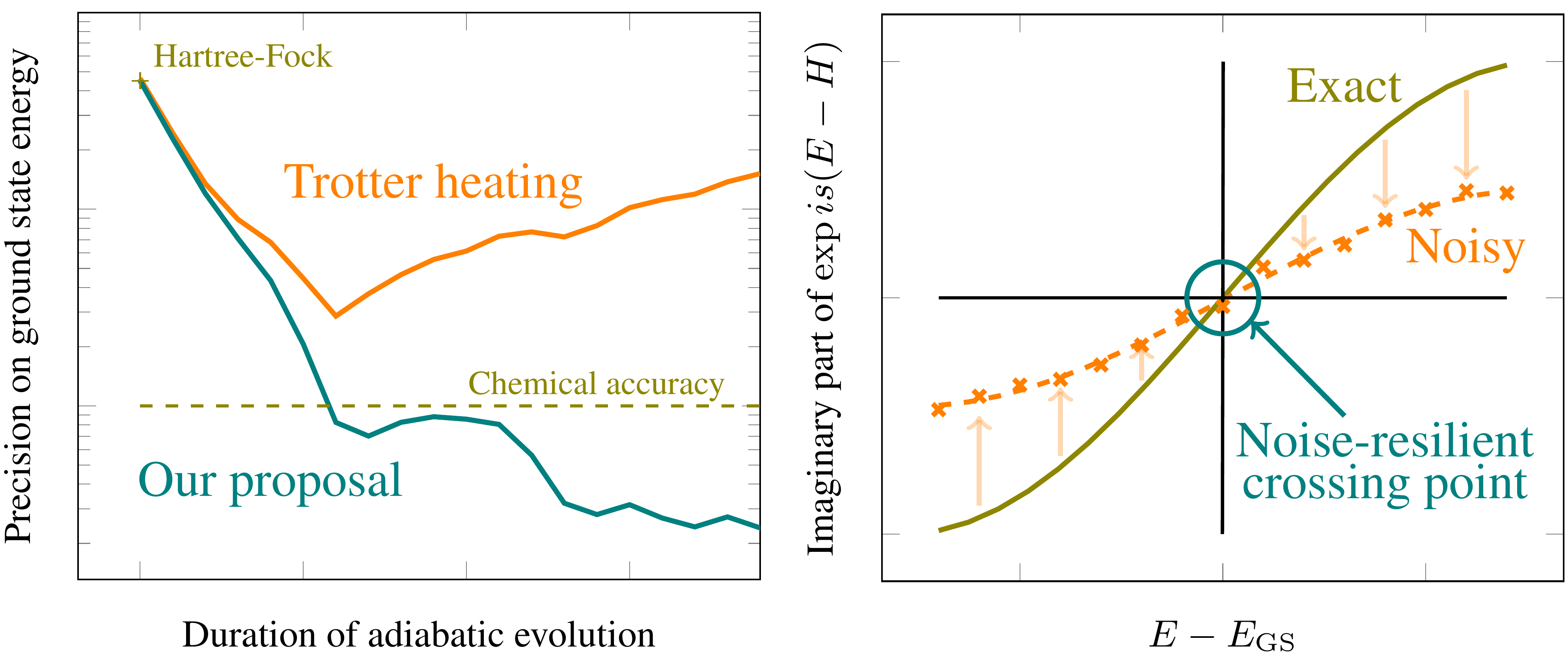}
    \caption{Conceptual figure summarizing our proposal. 
\emph{Left panel:} accuracy on the energy of the ground state prepared adiabatically, as a function of the total adiabatic time evolution. A Trotter decomposition leads to heating, precluding reaching chemical accuracy. Our proposal implements the exact adiabatic evolution and is free of any discretization error heating. \emph{Right panel:} imaginary part of $e^{is(E-H)}$ on the state prepared, as a function of $E$, at fixed arbitrary $s$. The main effect of hardware noise is to flatten the curve, which does not affect the point where the curve vanishes, that is the ground state energy $E_{\rm GS}$. The data is obtained from an H${}_6$ hydrogen chain for the left panel and H${}_2$ hydrogen molecule for the right panel, both in a STO-3G basis, see Section \ref{resourcesection} and below.}
\label{conceptualfigure}
\end{figure}

Recently, a new randomised ``TETRIS" algorithm (Time Evolution Through Random Independent Sampling) for Hamiltonian simulation with the following features was proposed~\cite{granet2023continuous}: (i) the cost depends not on the number of terms in the Hamiltonian but on the interaction norm $\mu_I$ (which we will introduce shortly) which is bounded by the 1-norm and generically has a scaling between $O(L)$ and $O(L^3)$ depending on the system and the decomposition considered~\cite{rubin2018application,lee2021even,koridon2021orbital}, (ii) the number of gates can be chosen at will at the cost of introducing sampling overhead, (iii) time-dependent Hamiltonians can be simulated at no extra cost, (iv) there are no parameters to be optimised, (v) the output is free from discretization (Trotter) error and (vi) in practice, the required number of gates not only scales advantageously but also comes with a prefactor known to be $<1$.

This algorithm removes thus one of the main obstacles to adiabatic state preparation (ASP) for chemistry applications and, indeed, our first main contribution is to show that states that hold information about the ground state energy of small molecules within chemical accuracy can be prepared adiabatically in this way with circuits containing only tens to hundreds of two-qubit gates. For example, with around $\sim 10^3$ two-qubit gates, one can prepare the ground state of molecules like LiH or a hydrogen chain H${}_4$ at equilibrium within chemical precision. 

State preparation is only useful to the extent that observables (most notably the energy) can be extracted from the state. In chemistry systems, measuring the energy is itself a non-trivial task, both because of (i) the high number of terms in the Hamiltonian and (ii) the high precision required. Multiple techniques developed for VQE solved the first problem and allow for optimal measurement of the Hamiltonian terms, see e.g. \cite{tilly2022variational}. However, they do not remove the high precision obstacle.
Our second main contribution is to overcome this challenge by developing three approaches to efficiently measure arbitrary observables in a state prepared by the TETRIS method. The measurement techniques are resilient to noise and we show that the combined programme of adiabatic state preparation plus measurement yields chemically accurate results on a 4-qubit problem using 100'000 $<(1\mathrm{e}{-3})^{-2}$ shots on a noisy quantum computer with errors modeled by depolarising noise on two-qubit gates with fidelity 99.84\% \cite{moses2023racetrack,decross2024computational} and \emph{without error mitigation} apart from trivial symmetry filtering.

To summarise, the two-step method we propose is depicted in Fig. 1 and consists of:

\begin{itemize}
    \item an \emph{adiabatic preparation} of the ground state $\mathcal{A}|{\rm ini}\rangle$ where $|{\rm ini}\rangle$ is some initial state and $\mathcal{A}$ the operator implementing the adiabatic path. Using the algorithm of \cite{granet2023continuous}, this step can be implemented without any discretization error heating.
    \item an \emph{energy measurement} to determine the energy of the state $\mathcal{A}|{\rm ini}\rangle$. We propose three different related methods: a binary search approach, an arctan fit approach, and a Robbins-Monro approach. These three methods possess some built-in noise resilience.
\end{itemize}
The paper is organized as follows. We start in Section \ref{setup} with a quick description of the setup encountered in quantum chemistry applications of quantum computing. In Section \ref{adiabatic}, we present the algorithm of \cite{granet2023continuous} to implement time evolution under a time-dependent Hamiltonian. We discuss how to best choose the parameters of the algorithm and give some runtime estimate for ASP. In Section \ref{measuringenergy} we then discuss how to best measure the energy of the state that we prepared adiabatically. We introduce three different approaches for the computation of the energy. Then in Section \ref{resourcesection} we do a resource estimate of the algorithm to reach chemical precision in multiple chemical systems. In particular we investigate the scaling with system size of the time that is required to perform ASP. Then in Section \ref{simulationssections} we present some noisy simulations of the algorithm. Finally we summarize in Section \ref{conclusion} the findings and conclusions of our paper.

\section{II. Setup and notations}
\setlabel{II}{setup}
Quantum chemistry aims at computing the ground state energy of molecules \cite{jensen2017introduction}. These are described by the following Hamiltonian
\begin{equation}
\begin{aligned}
    H=&-\sum_i \frac{\partial_{R_i}^2}{2M_i}-\sum_{j}\frac{\partial_{r_j}^2}{2}-\sum_{i,j}\frac{Z_i}{|R_i-r_j|}\\
    &+\sum_{i_1<i_2}\frac{Z_{i_1}Z_{i_2}}{|R_{i_1}-R_{i_2}|}+\sum_{j_1<j_2}\frac{1}{|r_{j_1}-r_{j_2}|}\,,
\end{aligned}
\end{equation}
where $R_i,r_j,M_i,Z_i$ are respectively the positions of the nuclei, the positions of the electrons, the masses of the nuclei and the charges of the nuclei, written here in atomic units. The movement of the nuclei can usually be neglected with the Born-Oppenheimer approximation \cite{jensen2017introduction}, leaving only electronic degrees of freedom $r_j$ and a parametric dependence of the ground state energy in the nuclei positions $R_i$. The Hilbert space associated to this Hamiltonian is infinite-dimensional, and has to be discretized in some way to be implemented on a computer (whether classical or quantum). Approaches referred to as ``first quantization" discretize space \cite{babbush2018low,babbush2019quantum,su2021fault}, while  ``second quantization" methods project the Hilbert space onto linear combinations of a finite number of well-chosen states that approximate equilibrium configurations, called basis set. In this paper we will consider the second quantization approach, which is the most studied. Here, the Hamiltonian is (approximately) written as
\begin{equation}
H=\sum_{p,q=1}^Lh_{pq}c^\dagger_pc_q+\sum_{p,q,r,s=1}^Lh_{pqrs}c^\dagger_pc^\dagger_qc_rc_s\,,
\end{equation}
where $L$ is the number of orbitals (which depends on the basis set), $h_{pq},h_{pqrs}$ are real coefficients, and $c_p,c^\dagger_p$ are fermionic operators satisfying canonical anticommutation relations $\{c_p,c^\dagger_q\}=\delta_{p,q}$ and $\{c_p,c_q\}=0$. To be implemented on a quantum computer, this second-quantized Hamiltonian must be written in terms of Pauli matrices $X,Y,Z$. This can be done through different mappings \cite{jordan1993paulische,bravyi2002fermionic,seeley2012bravyi}. We will consider the widely used Jordan-Wigner transformation
\begin{equation}
    c_p=\frac{1}{2}(X_p+iY_p)Z_{p-1}Z_{p-2}...Z_1\,.
\end{equation}
Once the Hamiltonian $H$ is written in terms of Pauli matrices, our objective is to determine its ground state with some precision $\epsilon$. In chemistry applications, this precision is very often set to $1.6\cdot 10^{-3}=1.6$mH, called chemical accuracy, which corresponds approximately to the accuracy achieved in experiments \cite{peterson2012chemical}. We will take the precision to be $10^{-3}$ for simplicity. We note that in this paper we will not be concerned with the accuracy of the basis set:  the precision we wish to achieve is with respect to the ground state of the (approximated, truncated) Hamiltonian $H$ expressed in a finite basis set, not with respect to original infinite-dimensional Hamiltonian. This is referred to as chemical \emph{precision} \cite{elfving2020will}.

\section{III. Adiabatic preparation}
\setlabel{III}{adiabatic}

\subsection{A. Adiabatic path}
In this Section, we present the first step of our proposal, which is the adiabatic preparation of the ground state of $H$. ASP relies on the \emph{adiabatic theorem} of quantum mechanics, which roughly states that if parameters of the Hamiltonian are varied slowly enough, a system initially in the ground state will remain in the ground state of the time-dependent Hamiltonian at all times \cite{born1928beweis,kato1950adiabatic,albash2018adiabatic}. Specifically, one considers a time-dependent Hamiltonian $H(u)$ for $0\leq u\leq 1$ and an initial state $|{\rm ini}\rangle$ such that (i) $|{\rm ini}\rangle$ is the ground state of $H(0)$, (ii) the final time Hamiltonian $H(1)$ is the target Hamiltonian $H$, and (iii) for all $u$, $H(u)$ is gapped. Then, evolving $|{\rm ini}\rangle$ with the time-dependent Hamiltonian $H(t/T)$ from $t=0$ to $t=T$, one prepares a state $\mathcal{A}(T)|{\rm ini}\rangle$ that becomes closer to the ground state of $H$ as $T$ grows larger. The convergence of the energy of $\mathcal{A}(T)|{\rm ini}\rangle$ is typically polynomial in $1/T$, but can be exponential in $T$ if the path is smooth enough \cite{jansen2007bounds,cheung2011improved,mackenzie2006perturbative}. The time needed then scales as the inverse square of the smallest gap of $H(u)$ along the path \cite{elgart2012note}. The operator $\mathcal{A}(T)$ that implements this time evolution under a time-dependent Hamiltonian can be written with the following time-ordered exponential
\begin{equation}
    \mathcal{A}(T)=\mathcal{T}\exp\left(i\int_0^T H(\tfrac{t}{T})\D{t}  \right)\,.
\end{equation}
There are obviously infinitely many valid paths $H(u)$ that can be considered for ASP. In our case, we will assume the simple following setting that is well adapted to quantum chemistry Hamiltonians. We split $H$ into two terms
\begin{equation}\label{hbi}
    H=H_B+H_I\,,
\end{equation}
where the ground state of the \emph{background} Hamiltonian $H_B$ contains only $Z$ Pauli matrix terms. More generally, one could consider any Hamiltonian $H_B$ such that its ground state is easy to prepare on a quantum computer. We decompose then the \emph{interaction} Hamiltonian $H_I$ that contains the remaining terms into Pauli strings
\begin{equation}
    H_I=\sum_{n=1}^{N_I}c_nP_n\,,
\end{equation}
with $P_n\in \pm\{ I,X,Y,Z\}^{\otimes L}$ and $c_n$ real coefficients. Up to flipping $P_n$ into $-P_n$, we can assume $c_n\geq 0$. For notational convenience, we will denote as well
\begin{equation}
    H_B=\sum_{n=N_I+1}^N c_nP_n\,,
\end{equation}
where here $P_n$ are strings of $Z$ Pauli matrices at some sites times a $\pm$ sign, and with $N$ the total number of terms in the Hamiltonian $H$. Then we take the initial state $|{\rm ini}\rangle$ as the ground state of $H_B$, which can always be prepared efficiently on a quantum computer as it is a product state. For a given weight function (or path) $w(u)\geq 0$ such that $w(0)=0$ and $w(1)=1$, we then set
\begin{equation}
    H(u)=H_B+w(u)H_I\,.
\end{equation}
This adiabatic path could be refined by e.g. setting a different weight $w(u)$ for every Pauli string in $H_I$. For simplicity, we will not do this and only consider a single weight function for all terms throughout this work. 

\subsection{B. Algorithm for implementing \texorpdfstring{$\mathcal{A}(T)$}{Lg}}
\setlabel{III. B}{algorithm0}
The operator $\mathcal{A}(T)$ can be implemented \emph{exactly} with a finite circuit depth on a quantum computer, using the  randomization algorithm introduced in \cite{granet2023continuous}. To present this algorithm, we define
\begin{equation}
    z(u)=\int_0^u w(u')\D{u'}\,,
\end{equation}
as well as $z^{-1}(u)$ the reciprocal function of the increasing function $z(u)$, namely such that $z^{-1}(z(u))=u$ for all $u$. We will denote
\begin{equation}
    \zeta=z(1)=\int_0^1w(u')\D{u'}\,,
\end{equation}
which is of order $\mathcal{O}(1)$. The algorithm takes as a parameter a gate angle $0<\tau<\pi/2$. It consists of the following steps.
\begin{enumerate}
\item Initialize the state of the system $|\psi\rangle$ to $|{\rm ini}\rangle$.
    \item For $n=1,...,N_I$, draw an integer $m_n\geq 0$ from a Poisson distribution with parameter $\tfrac{c_n\zeta T}{\sin\tau}$. 
    \item For $n=1,...,N_I$, draw $m_n$ real numbers $\tilde{t}_{i,n}$ (with $i=1,...,m_n$), independently and uniformly at random between $0$ and $\zeta$. For every $i,n$, set $t_{i,n}=T\, z^{-1}(\tilde{t}_{i,n})$.
    \item For $n=N_I+1,...,N$, draw an integer $m_n\geq 0$ from a Poisson distribution with parameter $\tfrac{c_n T}{\sin\tau}$.
    \item For $n=N_I+1,...,N$, draw $m_n$ real numbers $t_{i,n}$ (with $i=1,...,m_n$), independently and uniformly at random between $0$ and $T$.
    \item Find the sequence $(i_1,n_1),...,(i_M,n_M)$ such that
    \begin{equation}
        0<t_{i_1,n_1}<...<t_{i_M,n_M}<T\,,
    \end{equation}
    where $M=\sum_{n=1}^{N}m_n$.
    \item For $m=1,...,M$, apply the operator
    \begin{equation}
        \exp\left(i\tau P_{n_m}\right)
    \end{equation}
    on the state $|\psi\rangle$.
\end{enumerate}

Let us denote $U$ the random unitary operator that is generated with this process. The result of \cite{granet2023continuous} is that
\begin{equation}
    \mathbb{E}[U]=e^{-\tan(\tau/2)T( \zeta \mu_I+\mu_B) } \mathcal{A}(T)\,,
\end{equation}
where $\mathbb{E}$ denotes the statistical average with respect to $U$, and with $\mu_I,\mu_B$ the $1$-norm of the interaction/background Hamiltonian
\begin{equation}
    \mu_I=\sum_{n=1}^{N_I}c_n\,,\qquad\mu_B=\sum_{n=N_I+1}^{N}c_n\,.
\end{equation}
We will call \emph{attenuation factor} the term
\begin{equation}
    \lambda=e^{-\tan(\tau/2)T( \zeta \mu_I+\mu_B) }\,.
\end{equation}
In each random realization $U$, there are in average $\frac{c_n\zeta T}{\sin\tau}$ gates $e^{i\tau P_n}$ in the circuit for $n=1,...,N_I$, and $\frac{c_n T}{\sin\tau}$ gates $e^{i\tau P_n}$ for $n=N_I+1,...,N$. Let us denote $g_n$ the number of two-qubit gates that is required to implement $e^{i\tau P_n}$. Then the average number of two-qubit gates in the circuit is
\begin{equation}
    N_{TQG}=\frac{ T }{\sin\tau}\left(\zeta\sum_{n=1}^{N_I}g_n c_n+\sum_{n=N_I+1}^N g_nc_n\right)\,,
\end{equation}
which is finite even though the exact adiabatic preparation is implemented on average.

\subsection{C. Improvement of the algorithm for implementing \texorpdfstring{$\mathcal{A}(T)$}{Lg}}
\setlabel{III. C}{algorithm}
The algorithm of Section \ref{algorithm0} possesses the following improvement in the case where the time evolution with respect to the background Hamiltonian $H_B$ can be implemented with just a finite number of gates, namely when $e^{i\tau H_B}$ can be decomposed exactly as a finite product of gates on the quantum computer. This improvement is called the \emph{background Hamiltonian technique} \cite{granet2023continuous}. It consists in applying continuously the time evolution with respect to $H_B$, and randomizing only the time evolution with respect to the interaction Hamiltonian $H_I$. The upside is that the attenuation factor $\lambda$ involves only $\mu_I$ the $1$-norm of the interaction Hamiltonian. 

This improved algorithm takes as a parameter a gate angle $0<\tau<\pi/2$, and consists of the following steps:
\begin{enumerate}
\item Initialize the state of the system $|\psi\rangle$ to $|{\rm ini}\rangle$.
    \item For $n=1,...,N_I$, draw an integer $m_n\geq 0$ from a Poisson distribution with parameter $\tfrac{c_n\zeta T}{\sin\tau}$.
    \item For $n=1,...,N_I$, draw $m_n$ real numbers $\tilde{t}_{i,n}$ (with $i=1,...,m_n$), independently and uniformly at random between $0$ and $\zeta$. For every $i,n$, set $t_{i,n}=T\, z^{-1}(\tilde{t}_{i,n})$.
    \item Find the sequence $(i_1,n_1),...,(i_M,n_M)$ such that
    \begin{equation}
        0<t_{i_1,n_1}<...<t_{i_M,n_M}<T\,,
    \end{equation}
    where $M=\sum_{n=1}^{N_I}m_n$.
    \item For $m=1,...,M$, apply the operator
    \begin{equation}
        \exp\left(i\tau P_{n_m}\right) \exp\left(i(t_{i_m,n_m}-t_{i_{m-1},n_{m-1}}) H_B\right)
    \end{equation}
    on the state $|\psi\rangle$, with $t_{i_0,n_0}\equiv 0$.
    \item Apply $e^{i(T-t_{i_M,n_M})H_B}$ on the state $|\psi\rangle$.
\end{enumerate}

Let us denote $U$ the random unitary operator that is generated with this process. We have
\begin{equation}
    \mathbb{E}[U]=e^{-\tan(\tau/2) \zeta T\mu_I } \mathcal{A}(T)\,,
\end{equation}
where $\mathbb{E}$ denotes the statistical average with respect to $U$. In that case, we will call attenuation factor the term
\begin{equation}
    \lambda=e^{-\tan(\tau/2) \zeta T\mu_I }\,.
\end{equation}
In each random realization $U$, there are in average $\frac{c_n\zeta T}{\sin\tau}$ gates $e^{i\tau P_n}$ in the circuit. Let us denote $g_n$ the number of two-qubit gates that is required to implement $e^{i\tau P_n}$. Then the average number of two-qubit gates in the circuit is
\begin{equation}
    N_{TQG}=\frac{\zeta T \mu_I g}{\sin\tau}\,,\qquad g=\frac{\sum_{n=1}^{N_I}c_ng_n}{\sum_{n=1}^{N_I}c_n}\,,
\end{equation}
which is again finite.

\subsection{D. Choice of gate angle}
\subsubsection{1. Optimal gate angle on a noiseless quantum computer}

\begin{figure*}
\begin{center}
\includegraphics[scale=0.23]{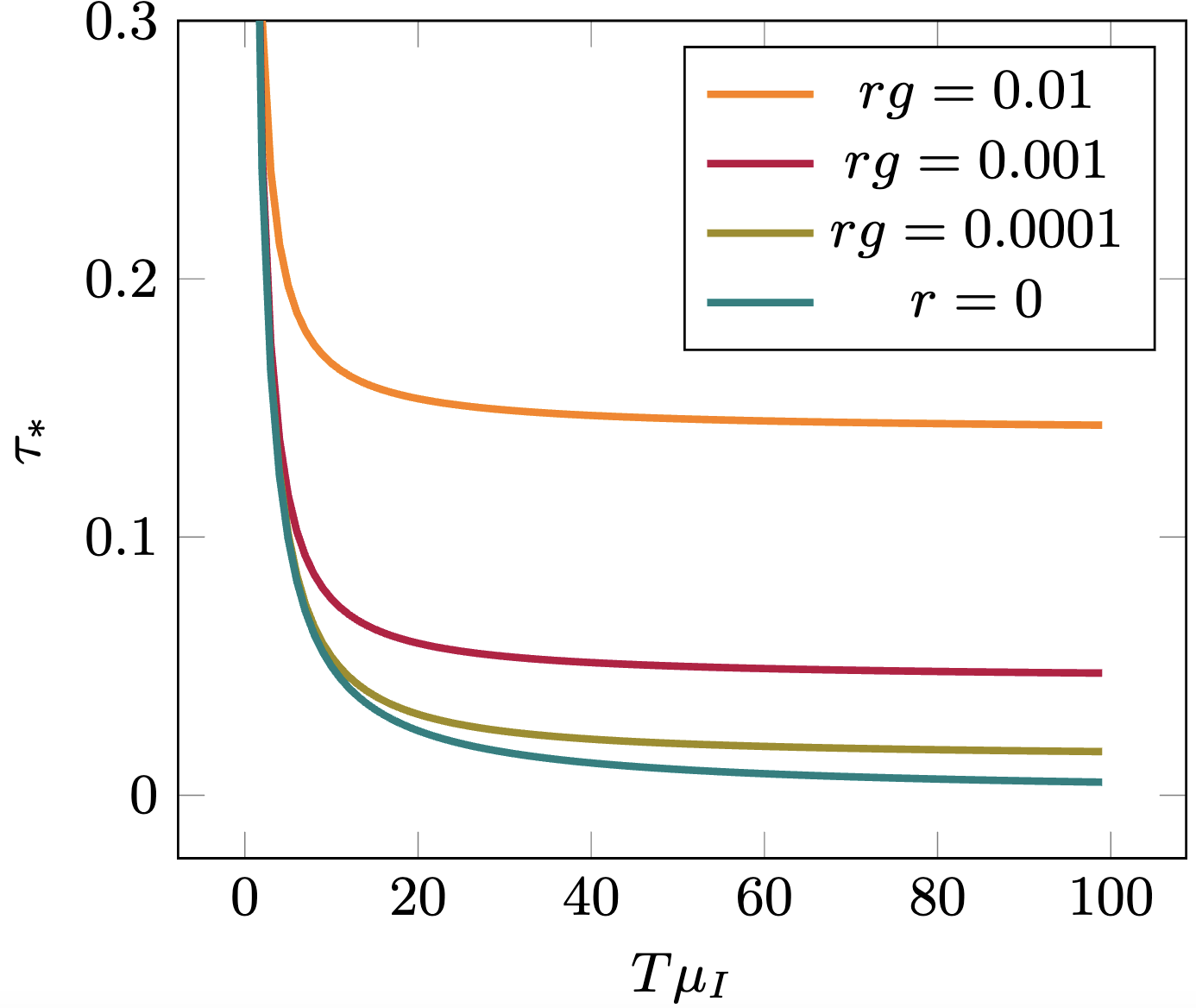}
$\qquad\qquad$
\includegraphics[scale=0.23]{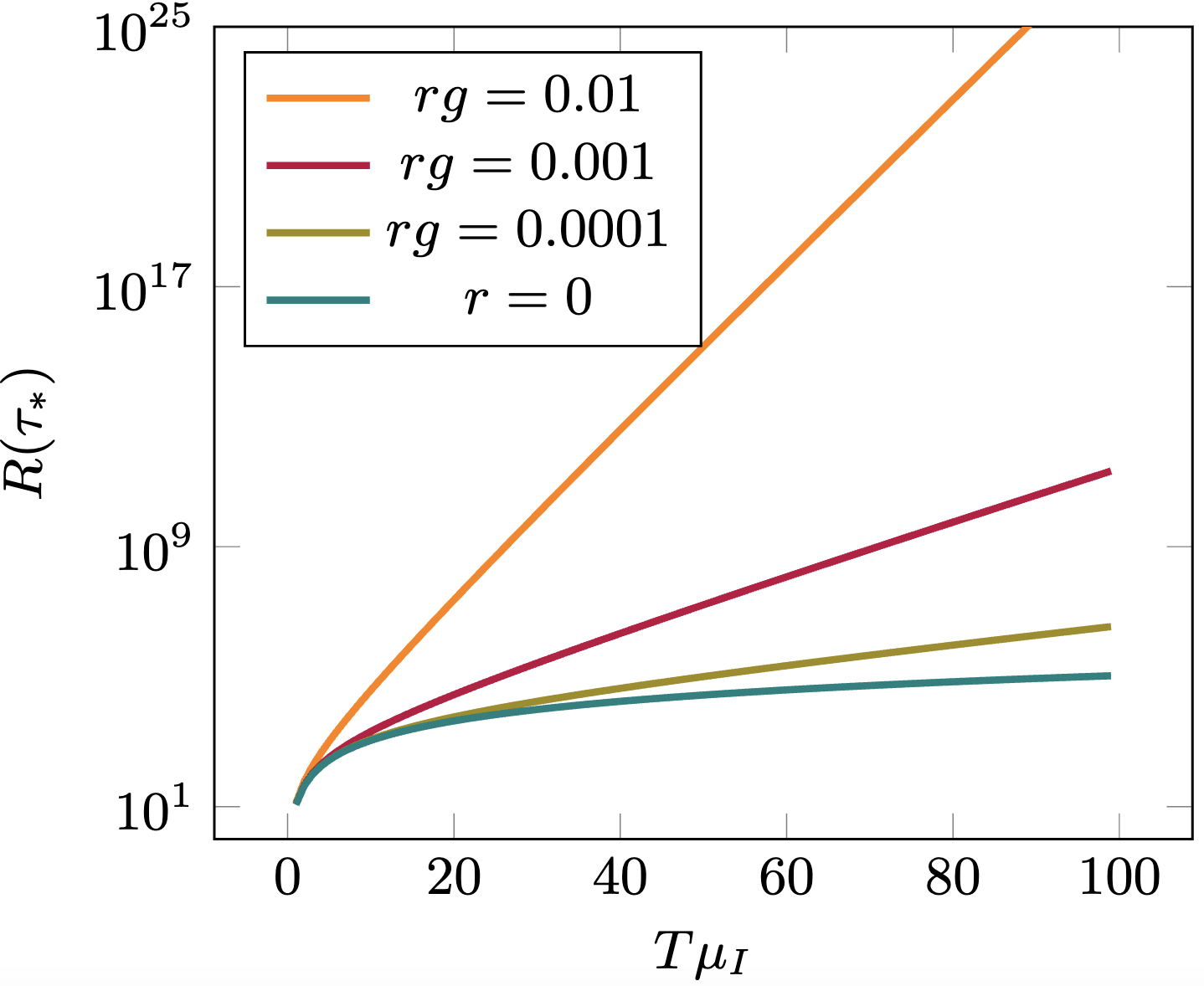}

\caption{Optimal gate angle $\tau_*$ (left) and minimal total runtime $R(\tau_*)$ (right) as a function of $T\mu_I$, with $\zeta=1$, for different values of noise $rg$. Total runtime means number of shots times number of two-qubit gates per shot to get a precision $\mathcal{O}(1)$ on the result. The total runtime to get a precision $\mathcal{O}(\epsilon)$ on the result is $R(\tau)/\epsilon^2$.}
\label {gateangles}
\end{center}
\end {figure*}

The gate angle $\tau$ is here a free parameter of the algorithm, that does not change the precision on the result. Increasing $\tau$ reduces the number of two-qubit gates $N_{TQG}$, so reduces the runtime of every random circuit. However, it also makes the attenuation factor $\lambda$ smaller. This attenuation factor increases the number of shots required to reach a certain precision. Indeed, because we have to divide the amplitude measured on the quantum computer by $\lambda$, the number of shots required to have a precision $\epsilon$ on the result is $(\epsilon\lambda)^{-2}$. We elaborate on this fact in Section \ref{samplingcosts} below. This means that for the same precision we have to run more shots of the same circuit as $\tau$ increases. Hence, the \emph{total runtime} at fixed precision, i.e. the product of runtime of each circuit and number of shots per circuit, has a non-trivial behaviour with the gate angle $\tau$ and reaches a minimum at some optimal $\tau_*$. This optimal gate angle $\tau_*$ moreover depends on the task that we wish to fulfil. Let us consider for example the problem of measuring an observable $\mathcal{O}$ within the adiabatically evolved state $\mathcal{A}(T)|{\rm ini}\rangle$. Using the above protocol, we can write
\begin{equation}
    \langle {\rm ini}|\mathcal{A}(T)^\dagger \mathcal{O} \mathcal{A}(T)|{\rm ini}\rangle=\lambda^{-2}\mathbb{E}_{U_1,U_2}[\langle {\rm ini}|U_2^\dagger\mathcal{O} U_1|{\rm ini}\rangle]\,,
\end{equation}
where the circuits $U_1,U_2$ are drawn independently from the previous protocol. Importantly, the backward and forward propagations that are randomly drawn do not (at least not necessarily) coincide $U_1\neq U_2$, so the amplitude $\langle {\rm ini}|U_2^\dagger\mathcal{O} U_1|{\rm ini}\rangle$ has to be measured explicitly on the quantum computer with for example a Hadamard test. This means that both $U_1$ and $U_2$ have to be implemented explicitly, so that the total number of two-qubit gates (not taking into account those possibly entering $\mathcal{O}$) is $2N_{TQG}$. Each of the two averages over $U_1,U_2$ contributes with an attenuation factor $\lambda$, so that the total attenuation is $\lambda^2$. On a perfect quantum computer, one thus has to run $\lambda^{-4}$ many circuits to obtain a precision $\mathcal{O}(1)$ on the result, and $\lambda^{-4}\epsilon^{-2}$ to obtain a precision $\epsilon$. Hence the total number of two-qubit gates to run (across different circuits) to get precision $\epsilon$ is $R(\tau)/\epsilon^2$, with $R(\tau)=2N_{TQG}\lambda^{-4}$.  
Explicitly, this factor $R(\tau)$ is
\begin{equation}
     R(\tau)=\frac{2\zeta T \mu_I g}{\sin \tau}\exp\left(4\tan(\tau/2)\zeta T\mu_I\right)\,.
\end{equation}
We recall that $R(\tau)/\epsilon^2$ is the \emph{total} number of two-qubit gates to run to reach precision $\epsilon$, but each circuit has only $2N_{TQG}$ two-qubit gates, independently of $\epsilon$.

This total runtime reaches a minimum at some $\tau_*$. Imposing $\partial_\tau R(\tau_*)=0$, we find  at large $T\mu_I$ \cite{granet2023continuous}
\begin{equation}
    \tau_*= \frac{1}{2\zeta T\mu_I}\,,\qquad R(\tau_*)=\mathcal{O}(T^2\mu_I^2)\,.
\end{equation}
Intuitively, in absence of noise, one should choose the largest possible $\tau$ to reduce the number of two-qubit gates in each circuit, while still keeping an attenuation factor $\lambda$ of order $\mathcal{O}(1)$. This gives $\tau=\mathcal{O}(1/( T \mu_I))$.

\subsubsection{2. Optimal gate angle on a noisy quantum computer}
\setlabel{III. D. 2}{optimalnoise}
This ``total runtime" count holds for a perfect, noiseless quantum computer where every gate can be executed perfectly. However, in any hardware in the near future, noise cannot be neglected and is a crucial aspect to take into account in quantum computing algorithms. We are going to make the assumption that because of the imperfect hardware, the measured expectation value is multiplied by $e^{-r}$ with $r>0$ for every two-qubit gate present in the circuit. This $e^{-r}$ is called the effective two-qubit gate fidelity. This noise model is of course a crude approximation of actual hardware noise, which is much more complex and platform-dependent. However, it is known to give a rough first estimate \cite{moses2023racetrack} and there are several noise mitigation techniques such as Probabilistic Error Cancellation that can convert any noise occurring in the hardware into this kind of global depolarizing noise \cite{temme2017error,self2022protecting}. 
We note that even with quantum error correction, the noise level $r$ would not be exactly equal to $0$, so this discussion is also relevant beyond the NISQ era. As long as $rg_n$ is small (if $rg_n$ is not small a single gate $e^{i\tau P_n}$ would be too noisy to implement), the noise can then be assumed to multiply the measured amplitude by a factor $\sim q= e^{-rN_{TQG}}$. This has the same effect as the attenuation factor $\lambda$ and increases the number of shots required to reach a given precision. Hence, the total runtime with a noise parameter $r$ is
\begin{equation}
    R(\tau)=\frac{2\zeta T \mu_I g}{\sin \tau}\exp\left(\frac{4r\zeta T \mu_I g}{\sin\tau}+4\tan(\tau/2)\zeta T\mu_I\right)\,.
\end{equation}
Let us write an equation for the gate angle $\tau_*$ that minimizes $R(\tau)$. The quantity $x=\tan(\tau_*/2)$ satisfies the quartic equation
\begin{equation}
    x^4(4+2rg)\zeta T\mu_I-x^3+4x^2\zeta T\mu_I+x-2r\zeta T\mu_I g=0\,.
\end{equation}
If we set $r=0$, at large $T\mu_I$ we recover the previous noiseless result. In the presence of noise $r>0$, the behaviour of $\tau_*$ at large $ T\mu_I$ is completely different. We find at large $T\mu_I$
\begin{equation}
    \tau_*=\sqrt{2rg}\,,\qquad R(\tau_*)=\mathcal{O}(e^{4\sqrt{2rg}\zeta T\mu_I})\,.
\end{equation}
In the presence of noise, the argument of the exponential appearing in $R(\tau)$ cannot be kept constant when $ T\mu_I\to\infty$. The optimal gate angle depends only on the noise factor $r$, and the runtime is exponential in $ T\mu_I$, with an exponent proportional to $\sqrt{r}$. 

In Figure \ref{gateangles}, we show the behaviour of the optimal gate angle $\tau_*$ and the corresponding minimal runtime $R(\tau_*)$ as a function of $T\mu_I$, for different amounts of noise $r$. To give concrete numbers, $r\approx 2\cdot 10^{-3}$ for current state-of-the-art two-qubit gates \cite{moses2023racetrack}, while the average number of two-qubit gates per rotation $g$ is roughly $g\approx L/2$ where $L$ is the number of qubits, see Figure \ref{reducednorms} below.

\subsection{E. Techniques for reducing the norm \texorpdfstring{$\mu_I$}{Lg}}
\setlabel{III. E}{symme}
In terms of system size, the cost of the algorithm depends only on the $1$-norm $\mu_I$ of the interaction Hamiltonian. It differs from a Trotter decomposition that depends on the number of terms in the Hamiltonian, and also from Linear Combination of Unitaries techniques, as well as other randomization techniques such as qDRIFT, which depend on the $1$-norm of the entire Hamiltonian. Contrary to the $2$-norm, the $1$-norm is basis-dependent. There can thus be decompositions of the same Hamiltonian that have different interaction norms $\mu_I$. Since the cost of implementing the time evolution for time $t$ is a function of $t\mu_I$, it is valuable to use a decomposition that reduces this norm $\mu_I$.

There exist different techniques to decrease the $1$-norm. In particular, in the context of chemistry applications, one can perform a unitary transform on the orbital basis that minimizes the $1$-norm, instead of minimizing the energy of the Hartee-Fock state \cite{koridon2021orbital}. This classical pre-processing of the Hamiltonian scales polynomially with the number of orbitals. According to \cite{koridon2021orbital}, it can for example modify the scaling with system size of the $1$-norm of the Hamiltonian corresponding to a hydrogen chain, which then scales as $\sim L^{1.4}$ with $L$ the number of hydrogen atoms, instead of $\sim L^{2.1}$ in the Hartree-Fock optimized basis.

Another cheap way of reducing the $1$-norm is to make use of exact symmetries in the Hamiltonian. It is well-known that symmetries can be used to reduce the number of qubits, with a process called tapering \cite{bravyi2017tapering}. However, it does not necessarily reduce the $1$-norm of the Hamiltonian, which is the only relevant system-size-related parameter in the TETRIS algorithm. Instead, given a conserved quantity $Q$, i.e. an operator that commutes with the Hamiltonian $[H,Q]=0$, one can always add a multiple of $Q$ to form a new Hamiltonian $H'$
\begin{equation}
    H'=H-\alpha Q\,,
\end{equation}
that has the same eigenstates as $H$ with eigenvalues trivially shifted by their eigenvalue on $-\alpha Q$. In the quantum chemistry application case, particle number $N_{\rm part}$ is a conserved quantity. We find that setting $Q=N_{\rm part}^2$  the square of the number of particles, and taking $\alpha$ so as to minimize $\mu_I$, can significantly reduce the norm $\mu_I$. When using the Jordan-Wigner transformation for the encoding, one finds in that case that the optimal choice of $\alpha$ is given by the median value of the coefficients $c_n$ in front of Pauli strings with exactly two $Z$'s. In the molecules that we consider below in this paper, we systematically use this trick to reduce the norm of the Hamiltonian.

\section{IV. Measuring the energy}
\setlabel{IV}{measuringenergy}
\subsection{A. Generalities}
The algorithm that we presented in Section \ref{adiabatic} enables us to prepare (an approximation of) the ground state of the Hamiltonian
\begin{equation}
    |\psi(T)\rangle=\mathcal{A}(T)|{\rm ini}\rangle\,.
\end{equation}
One now needs to decide how to best measure this state and retrieve information in order to fulfill a certain task. In our case, we are interested in evaluating the energy of the state that we prepared
\begin{equation}
    E(T)=\langle \psi(T)| H |\psi(T)\rangle\,.
\end{equation}
The simplest approach to compute $E(T)$ is to measure individually every Pauli string in the Hamiltonian $H$. A Pauli string with a small coefficient $c_n$ requires fewer shots to obtain a certain fixed precision on the total energy. At a fixed total number of shots, the optimal strategy is to spend a number of shots to measure $P_n$ that is proportional to $c_n$ \cite{wecker2015progress,koridon2021orbital}. This way, in order to have a precision $\epsilon$ on the result, the number of shots $M_S$ that is required is
\begin{equation}
    M_S=\frac{\mu^2}{\epsilon^2}\,,
\end{equation}
where $\mu$ is the $1$-norm of the total Hamiltonian
\begin{equation}
    \mu=\sum_{n=1}^Nc_n=\mu_I+\mu_B\,.
\end{equation}
This approach has two shortcomings. Firstly, it is very sensitive to noise on an imperfect hardware. With this approach, obtaining a precision of order $10^{-3}$ would inevitably require to bring the noise in the signal below $10^{-3}$. 
Secondly, with this approach, one necessarily computes \emph{all} the bits in the binary decomposition of the energy $E(T)$, including those that can be obtained classically by exact or approximate methods. To make a better use of scarce quantum computing resources, one would like to be able to compute only the most precise bits that are beyond reach of classical methods. Moreover, the noise will not affect these bits that are easy to compute classically, and will completely blur the bits that require quantum computation. 

\subsection{B. Measuring the energy through time evolution}
\setlabel{IV.B}{noiseresi}
\subsubsection{1. Recall: QPE techniques}
In order to devise a better method to compute the energy of the state $|\psi(T)\rangle$, let us briefly review  Quantum Phase Estimation (QPE) \cite{kitaev1995quantum,nielsen2010quantum}. QPE is an algorithm to measure the eigenvalue of an eigenvector $|\psi\rangle$ of a unitary operator $U$, that requires only $\mathcal{O}(\log 1/\epsilon)$ shots and a circuit depth $\mathcal{O}(1/\epsilon)$ to reach a precision $\epsilon$. In our case we could apply QPE to measure the energy of the state $|\psi\rangle$ assumed to be the ground state, by taking $U=e^{iH}$. The algorithm works as follows. It takes as input a unitary operator $U$, an eigenvector $|\psi\rangle$ and a target precision $\epsilon=2^{-n}$ for some integer $n$, and returns $n$ bits of the binary decomposition of $x$ where $e^{2i\pi x}$ is the eigenvalue of $U$. Firsty, it prepares $n$ ancilla qubits in the state $|+\rangle$. Then it performs sequentially $n$ controlled-$U^{2^m}$ operations between the system and the ancillas for $m=1,...,n$, followed by an inverse Quantum Fourier Transform (QFT) on the ancillas. Finally, measuring the ancillas exactly provides the bits of the binary decomposition of $x$.

What makes QPE able to yield a drastic reduction of the number of shots compared to measuring all the terms in $H$ is the time evolution. Measuring $e^{iH}$ does not have a notable precision change compared to measuring $H$. However, if we measure the eigenvalue of $e^{isH}$ instead of $e^{iH}$ for some $s>0$, we only need a precision $s\epsilon$ on the eigenvalue of $e^{isH}$ in order to have a precision $\epsilon$ on the eigenvalue of $e^{iH}$ (or more precisely on its value modulo $2\pi/s$). If one takes $s=1/\epsilon$, one only requires $\mathcal{O}(1)$ shots to reach a precision $\epsilon$. To be more concrete, we can describe the following ``iterative" version of QPE \cite{dobvsivcek2007arbitrary,chen2020quantum}. Let us assume that $0\leq x<1$ the eigenvalue of $H$ on $|\psi\rangle$ has an exact binary decomposition $x=0.b_1...b_n=\sum_{k=1}^n \frac{b_k}{2^k}$ with $n$ bits $b_i\in\{0,1\}$. If we measure
\begin{equation}\label{imagqpe}
    \Im \langle \psi(T)| e^{2i\pi 2^{n-1} (H-2^{-n-1})}|\psi(T)\rangle\,,
\end{equation}
we obtain a measurement outcome $+1$ with probability $1$ if $b_n=1$, and $-1$ with probability $1$ if $b_n=0$. Then, we shift $H'=H-\frac{b_n}{2^n}$. If we measure
\begin{equation}
    \Im \langle \psi(T)| e^{i2\pi 2^{n-2} (H'-2^{-n})}|\psi(T)\rangle\,,
\end{equation}
we obtain $+1$ with probability $1$ if $b_{n-1}=1$, and $-1$ if $b_{n-1}=0$. Proceeding repeatedly, we can measure all the bits $b_1,...,b_n$. The downside of this ``iterative" version of QPE is that one needs to have an exact eigenstate of $H$, and not an approximation, whereas QPE would succeed for general states with a probability proportional to the overlap of the state that we have with the target eigenstate. However, the upside of this approach is that one only requires to run $n$ separate circuits each containing only one controlled time evolution, instead of one circuit containing $n$ sequential controlled time evolutions \cite{dobvsivcek2007arbitrary}. It reduces thus the circuit depth, which is valuable for current and near-term devices where gates have a moderate level of noise.

\subsubsection{2. Binary search approach}
\setlabel{IV.B.2}{bisectionsection}
The previous ``iterative QPE" approach reduces the circuit depth compared to the original QPE. However, the times during which one needs to evolve the system are still large, of order $10^3$ to get chemical precision. On current-day devices, these very deep circuits are not feasible. In fact, the approach can be simplified even a bit further. When measuring \eqref{imagqpe} to determine the $n$-th bit of the energy, one does not necessarily need to do a time evolution up to time $2^{n-1}$. Assuming knowing the bits $b_1,...,b_{n-1}$, in order to determine the bit $b_n$, one only needs to time evolve sufficiently long and take enough shots to be able to distinguish whether the eigenvalue after removing the bits $b_1,...,b_{n-1}$ is \emph{below or above} $2^{-n-1}$. This observation suggests the following approach. Instead of seeking to measure $H$, we wish to answer the following question:
\begin{equation}\begin{aligned}\label{question}
    &\text{Given an energy }E\text{ and an eigenstate }|\psi\rangle,\\
    &\text{ is its eigenvalue below or above }E\text{?} 
\end{aligned}
\end{equation}
By performing a binary search, one only needs to answer $\mathcal{O}(\log 1/\epsilon)$ questions \eqref{question} to locate with precision $\epsilon$ the energy of the state $|\psi\rangle$. Moreover, each answer to \eqref{question} provides \emph{one bit} of information about the ground state. Contrary to measuring $H$ directly, here we can choose which bit of information we want to measure.

Each of these yes/no questions \eqref{question} can be answered by measuring the quantity
\begin{equation}\label{rhos}
   \rho(s)= \Im\langle \psi| e^{is(E-H)}|\psi\rangle\,.
\end{equation}
We will call the time $s$ ``central time", to distinguish it from the adiabatic preparation time $T$. Let us denote $\delta=E-E_{\rm GS}$ where $E_{\rm GS}$ is the energy of the state $|\psi\rangle$ that we are looking for. To answer question \eqref{question}, one only needs to decide whether $\rho(s)$ is positive or negative when $0<s<\frac{\pi}{|\delta|}$. The actual value of $\rho(s)$ beyond its sign is irrelevant. 

To that end, we wish to choose $s$ so as to maximize the measured value of $\rho(s)$. This would both (i) minimize the number of shots required to claim that $\rho(s)$ is either positive or negative, and (ii) maximize the resilience to noise, as a larger $\rho(s)$ would make it less likely to have its sign flipped because of noise. We will take into account the effect of noise in determining the optimal choice of the central time $s$. Under the same assumptions as in Section \ref{optimalnoise}, the measured $\rho_{\rm mes}(s)$ can be written as
\begin{equation}
    \rho_{\rm mes}(s)=C\sin(s\delta)e^{-su}\,,
\end{equation}
where
\begin{equation}
    C=\exp\left(-\frac{2r\zeta T \mu_I g}{\sin\tau}-2\tan(\tau/2)\zeta T\mu_I\right)
\end{equation}
is a $\tau$-dependent factor that takes into account the adiabatic preparation of $|\psi\rangle$, and where $u$ depends on the gate angle $\tau$ as
\begin{equation}\label{utau}
    u(\tau)=\frac{rg\mu_I}{\sin\tau}+\tan(\tau/2)\mu_I\,.
\end{equation}
The value of the central time $s_*$ that maximizes $\rho_{\rm mes}(s)$ is
\begin{equation}
    s_*=\frac{1}{\delta}\arctan \frac{\delta}{u}\,.
\end{equation}
Since $\delta$ is unknown, this optimal value cannot be chosen exactly. Computing the derivative of $s_*$ with respect to $\delta$, and using that $\arctan(x)\geq \frac{x}{1+x^2}$ for $x\geq 0$, we find that $s_*$ is a decreasing function of $|\delta|$. Its maximal value is $1/u(\tau)$. Moreover, within the binary search approach described above, after $m$ searches we have located the ground state with precision $2^{-m}$, so we know that $|\delta|\leq 2^{-m}$. Hence, assuming we have $|\delta|\leq \delta_{0}$ for some known initial guess $\delta_0>0$, we deduce the following bound
\begin{equation}
    \frac{1}{\delta_{0}}\arctan \frac{\delta_{0}}{u}\leq s_*\leq \frac{1}{u}\,,
\end{equation}
which depends only on known factors. In the following, we will set the central time equal to the lower bound
\begin{equation}\label{stau}
    s(\tau)= \frac{1}{\delta_{0}}\arctan \frac{\delta_{0}}{u(\tau)}\,.
\end{equation}
 The total number of two-qubit gates in the circuit is
\begin{equation}
    N_{TQG}=(2\zeta T+s)\frac{ \mu_I g}{\sin\tau}\,.
\end{equation}
Hence it follows that the total runtime $R^{Q}_{\delta,\delta_0}(\tau)$ to answer one question \eqref{question}, assuming we know a bound $|\delta|\leq\delta_0$, is
\begin{equation}\label{adiabaticscaling}
        R^Q_{\delta,\delta_0}(\tau)=\frac{s(\tau)+2\zeta T}{\sin^2[s(\tau)\delta]}\frac{\mu_Ig}{\sin\tau}\exp\left[2(s(\tau)+2\zeta T)u(\tau)\right]\,,
\end{equation}
with $s(\tau)$ given in \eqref{stau} and $u(\tau)$ in \eqref{utau}. We recall that this total runtime is the number of two-qubit gates per circuit times the number of circuits to run.

\begin{figure}
\begin{center}
\includegraphics[scale=0.23]{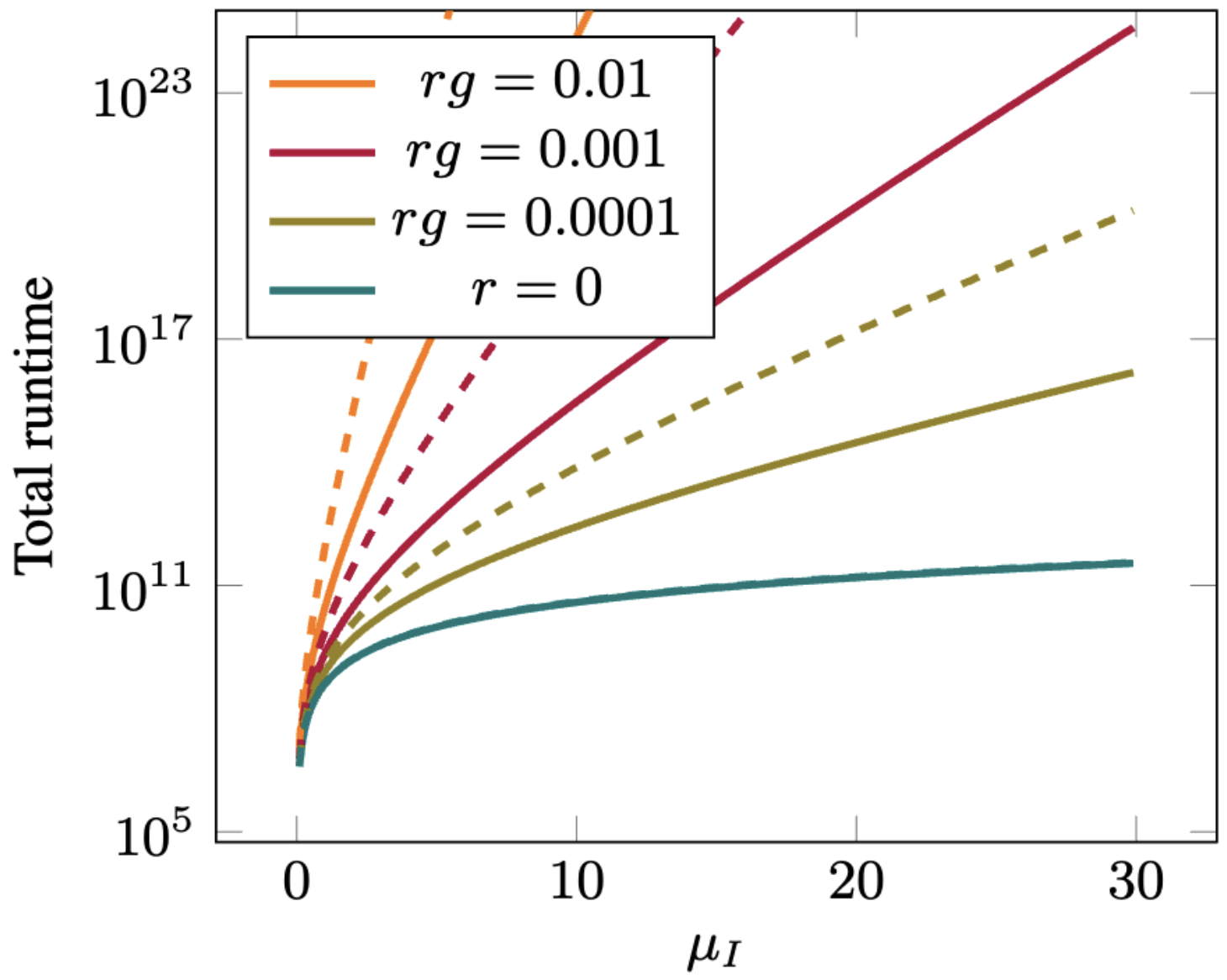}
\caption{Estimate of the minimal total runtime to get chemical precision with the binary search approach, as a function of $\mu_I$, with $\zeta=1/2$, $g=10$, for different values of noise $r$. Continuous lines are obtained with $T=10$ and dashed lines with $T=20$ (continuous and dashed lines are essentially superimposed for $r=0$). Total runtime is given in units of number of shots times number of two-qubit gates per shot, i.e. they are related to real runtimes by multiplying by the average duration of a two-qubit gate and dividing by the average number of gates that can are executed in parallel. The values plotted are upper bounds. As an example, the values of $\mu_I$ for chromium dimer in a STO-3G basis or FeMoCo with an active space of $50$ states are around $\mu_I\approx 10^3$.}
\label {totalruntime}
\end{center}
\end {figure}

In Figure \ref{totalruntime}, we plot the estimate of the total runtime to get chemical precision $1$mH with this binary search approach. Specifically, we plot the minimum with respect to the gate angle $\tau$ of $10R^Q_{\delta,\delta_0}(\tau)$ for $\delta=\delta_0=10^{-3}$, with $g=10$, $\zeta=1/2$, $T=10$, as a function of $\mu_I$. This value of $\delta$ corresponds to the last step of the binary search (when one already knows the location of the ground state within $2$mH), and the factor $10$ is a rough (over)estimate of the cost of previous steps. 
The analytical value of the total number of CNOTs to run to get chemical precision at large $\mu_IT$ on noiseless hardware is
\begin{equation}
   10 R^Q_{\delta,\delta}(\tau)=10eg\mu_I^2\left(\frac{\pi}{2\delta}+2\zeta T\right)^2\,,
\end{equation}
with $\delta=10^{-3}$. In Section \ref{simulationssections} below, we will present noisy implementations of this approach, and show that after simple noise mitigation the sign of the measured amplitude is indeed not affected by the noise.

\subsubsection{3. Arctan fit approach}
The approach we presented in Section \ref{bisectionsection} above spares circuit depth compared to the ``iterative" QPE by running a central time evolution $e^{is(H-E)}$ with $s$ just large enough so that the sign of the amplitude can be obtained. This allows one to answer the question \eqref{question}, namely whether the ground state energy is below or above $E$. We argued that this approach is noise-resilient because noise, while certainly affecting the measured value of $\rho(s)$ would be less likely to flip the sign of $\rho(s)$, especially when $s$ is large enough.

An improvement can be made further if the noise is mild enough to be described by a simple depolarizing channel. Namely, we make the following assumptions. Firstly, (i) we assume that the effect of the noise does not depend on $\delta$. This is a reasonable assumption as $\delta$ only enters as a phase in $\rho(s)$. If, for example, we compute $\rho(s)$ using a Hadamard test, the circuit depends on $\delta$ only through one single-qubit rotation. Secondly, (ii) we assume that the effect of the noise is just to multiply the measured $\rho(s)$ by some positive factor $q(s)$. This is a crucial assumption that means that noise only acts as a global depolarizing channel. We justified this assumption in Section \ref{optimalnoise} by pointing out that depolarising channels are the dominant sources of noise in many current hardware platforms and by referring to existing noise mitigation techniques that convert non-depolarizing noise into global depolarizing noise, like for example Probabilistic Error Cancellation \cite{temme2017error,self2022protecting}.

Here is the crucial consequence. Under these assumptions, although noise will affect individual amplitudes $\rho(s)$, \emph{it will not affect} the value of $E$ where $\rho(s)$ \emph{vanishes} (which is the ground state energy). We will test numerically this fact in Section \ref{noiseresinum} with noisy simulations. Specifically, let us assume we have an initial energy estimate $E_{\rm test}$ with an estimated precision $\epsilon>0$. We measure $\rho(s)$ for two different values of $E=E_{\rm test}\pm\epsilon$, but with the \emph{same} $s$, denoting $\rho_\pm$ the noisy measurement outcomes. Under these two assumptions, we can extract
\begin{equation}\label{formulalinear}
    E_{\rm GS}=E_{\rm test}+\frac{1}{s}\arctan \left(\tan(s\epsilon)\frac{\rho_++\rho_-}{\rho_--\rho_+}\right)\,.
\end{equation}
The advantage of this formula is that it depends on a \emph{ratio} of measurements of $\rho(s)$ at same value of $s$, and so would not be highly sensitive to depolarizing noise as it would cancel out from numerator and denominator. Using this formula to refine the estimate of the ground state energy is a way of utilising information in the amplitude of $\rho(s)$, which are, otherwise, thrown away in the binary search approach above (which only requires to know the sign of $\rho(s)$). We will present below in Section \ref{simulationssections} noisy numerical simulations of this approach and show that it performs well, showing that these assumptions are realistic.

\subsubsection{4. Robbins-Monro approach}
The previous ``arctan fit" approach relied on the observation that, in the presence of depolarizing noise, although the value of $\rho(s)$ is modified, the value of $E$ where $\rho(s)$ vanishes is invariant, which is $E_{\rm GS}$. We finally present a third approach to compute $E_{\rm GS}$ that uses an algorithm to find the zero of a function that we only know with statistical error.

The \emph{Robbins-Monro algorithm} is an algorithm for finding the root of a function $M(x)$ when instead of having directly access to $M(x)$, one has only access to a random variable $V(x)$ that averages to $M(x)$, i.e. $\mathbb{E}[V(x)]=M(x)$ \cite{robbins1951stochastic}. Specifically, the 
Robbins-Monro algorithm requires that (i) $M(x)$ is a non-decreasing function of $x$, (ii) $M'(x_*)>0$ where $M(x_*)=0$, and (iii) the values taken by the random variable $V(x)$ are uniformly bounded. In our case, this setup corresponds to the variable $x=E$ the trial energy, to the function $M(x)=\Im \langle \psi| e^{is(E-H)}|\psi\rangle$ at some fixed $s$, and with $V(x)$ the expectation value of one random circuit generated with the algorithm in Section \ref{algorithm}. The requirements of the Robbins-Monro algorithm are satisfied provided $E$ is in the interval $[E_{\rm GS}-\frac{\pi}{2s},E_{\rm GS}+\frac{\pi}{2s}]$. This algorithm takes as parameters an arbitrary sequence of steps $a_n>0$ such that
\begin{equation}
    \sum_{n=0}^\infty a_n=\infty\,,\qquad \sum_{n=0}^\infty a^2_n<\infty\,.
\end{equation}
Starting from an initial energy estimate $E_0$, we then build a sequence of energy estimates $E_n$ as
\begin{equation}\label{energyiterates}
    E_{n+1}=E_n-a_n V_n(E_n)\,,
\end{equation}
where $V_n(E_n)$ is the expectation value of random circuit generated with the algorithm in Section \ref{algorithm} to measure $\Im \langle \psi| e^{is(E_n-H)}|\psi\rangle$. This sequence of energies $E_n$ converges when $n\to\infty$ to the root of the function $\Im \langle \psi| e^{is(E-H)}|\psi\rangle$, namely the ground state energy $E_{\rm GS}$.

This approach shares with the previous arctan fit approach its noise resilience when the noise can be well approximated by a depolarizing channel, as it does not shift the value of $E$ where $\rho(s)$ vanishes. It has the advantage of being a well-studied algorithm with optimality results and improvements. However, the arctan fit approach above incorporates the known shape of the function $\rho(s)$ which might bring some additional precision. We will present in Section \ref{simulationssections} a comparison of these different approaches using a depolarizing channel.

\section{V. Resource estimate}
\setlabel{V}{resourcesection}
In this Section we report a resource estimate for obtaining chemical precision with the TETRIS method on different molecules.
\subsection{A. Hamiltonian-related costs}
\begin{figure*}
\begin{center}
\includegraphics[scale=0.23]{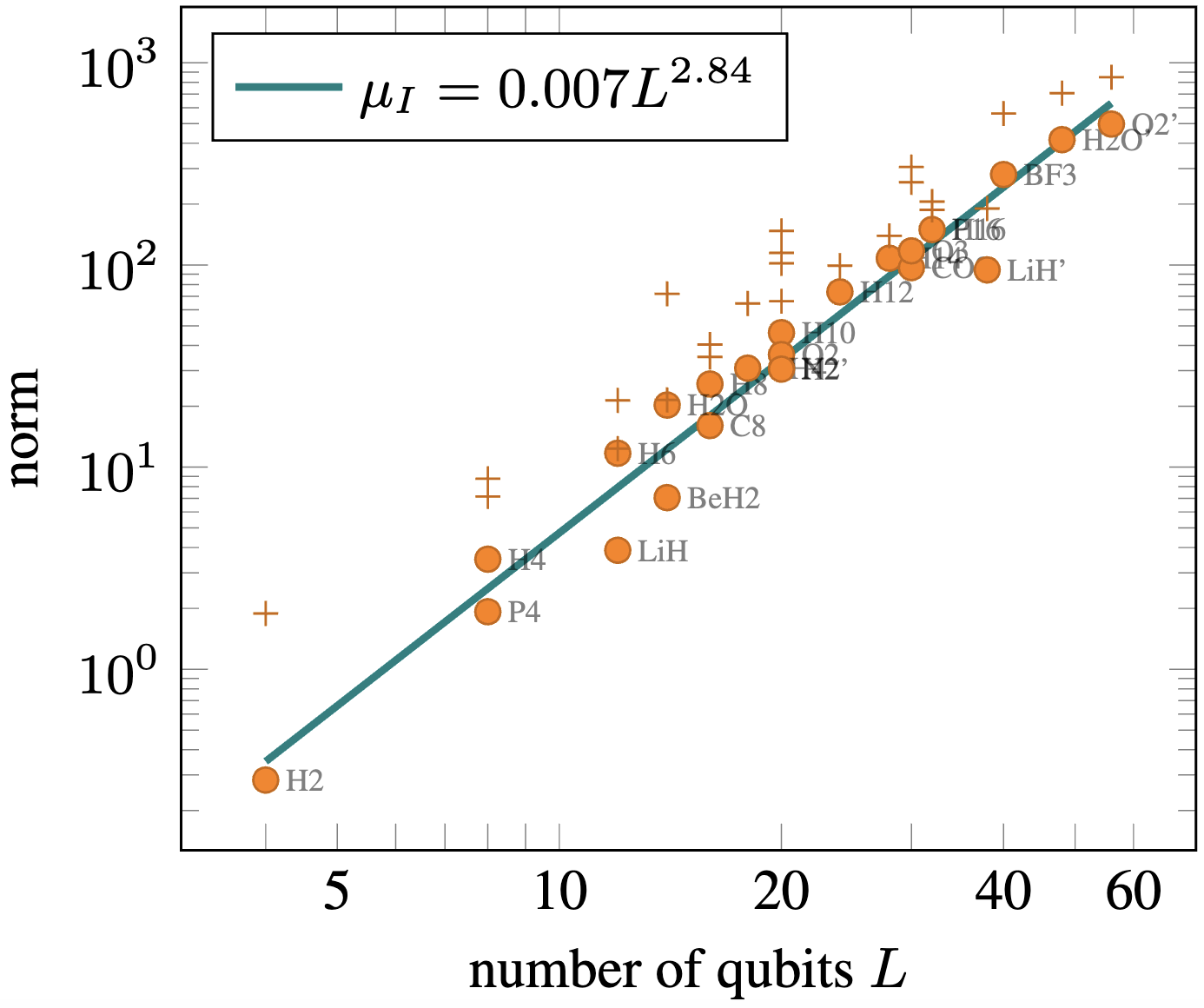}
$\qquad\qquad$
\includegraphics[scale=0.23]{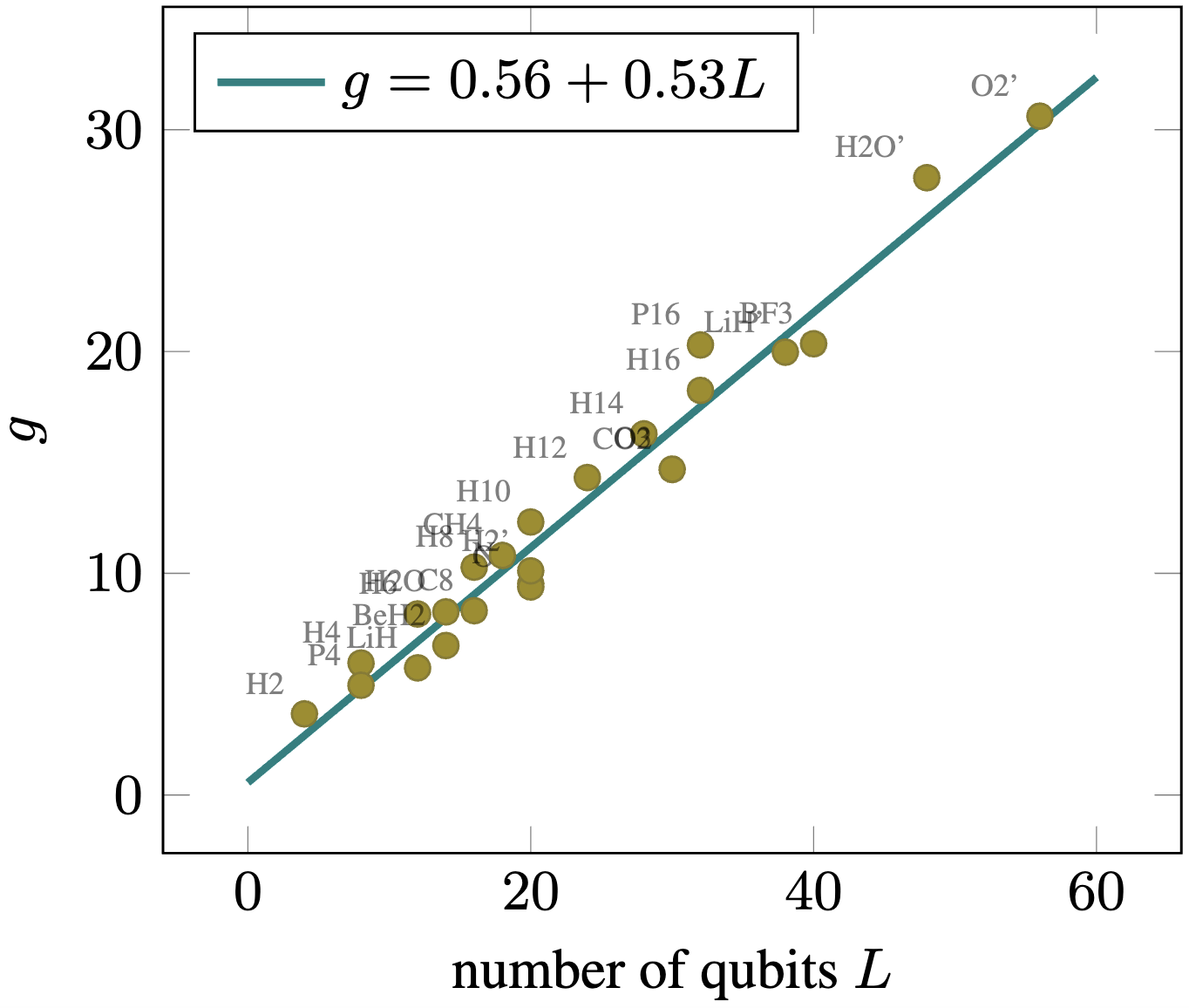}

\caption{Value of reduced norm $\mu_I$ (left panel, circles), total norm $\mu$ (left panel, crosses), factor $g$ assuming a Hadamard test and Jordan-Wigner decomposition (right panel), as a function of the number of qubits $L$.
The molecules considered are hydrogen chains H${}_{2n}$ for $n=1,...,8$, hydrogen squares P${}_{n^2}\equiv$ H${}_{n^2}$ for $n=2,4$, hydrogen cube Cub${}_{2^3}\equiv$ H${}_{2^3}$, N${}_2$, H${}_2$O, O${}_2$, LiH, BeH${}_2$, CO${}_2$, O${}_3$, BF${}_3$, CH${}_4$ in a STO-3G basis, and H${}_2$, H${}_2$O, LiH, O${}_2$, CH${}_4$ in a CCPVDZ basis (indicated by a $'$ on the plot), all in their equilibrium configuration.}
\label {reducednorms}
\end{center}
\end {figure*}
We start with a study of the required resources of the algorithm that pertain to the size of the molecular system studied. The number of qubits and the number of terms in the Hamiltonian $H$ do not enter directly the cost of the algorithm. The only dependence of the number of two-qubit gates in the system size is through the interaction norm $\mu_I$, which is the sum of the coefficients in the decomposition of the interaction Hamiltonian $H_I$ into Pauli strings, and through the factor $g$, which is the average number of two-qubit gates required to implement a rotation $e^{i\tau P_n}$.

In Figure \ref{reducednorms}, we study the dependence of these two parameters $\mu_I$ and $g$ on various molecules and basis sets as a function of the number of qubits (i.e. the number of spin orbitals). We decompose these molecular Hamiltonians in terms of Pauli strings, after applying a Hartree-Fock optimization of the molecular or spin orbitals. We call $\mu$ the $1$-norm of the Hamiltonians obtained this way, and $\mu_I$ the $1$-norm of the Hamiltonians without counting single Pauli $Z$ terms, see Section \ref{algorithm}, and after removing the square particle number operator to minimize the norm as explained in Section \ref{symme}. We observe that the norm $\mu_I$ scales polynomially with the number of qubits $L$, with a fit $\mu_I\approx 0.007 L^{2.84}$. 
When restricted to hydrogen chains, we observe a behaviour $\mu_I\approx 0.2 L^{2.13}$. We note that it is known that if instead of doing the Hartree-Fock optimization, we perform an orbital rotation to minimize the norm, then we can bring this exponent below $2$ \cite{koridon2021orbital}.
Moreover, we see that the norm $\mu_I$ of the interaction Hamiltonian is often a few factors below the norm of the entire Hamiltonian. The factor $g$ is seen to be well approximated by $g \approx L/2$.

\subsection{B. Adiabatic-path-related costs}
\setlabel{V.B}{adiabaticpathcosts}
\subsubsection{1. Minimal adiabatic time to reach chemical precision}
We now estimate the behaviour of the required adiabatic time evolution $T$ to reach chemical precision, as a function of the system size. We study two different paths that have different smoothness behaviour. The first one is linear
\begin{equation}\label{linearweight}
    w(u)=u\,,
\end{equation}
for which we have $\zeta=1/2$, and the second one has a quadratic behaviour both at $u=0$ and $u=1$
\begin{equation}\label{quadraweight}
    w(u)=2u^2-u^4\,,
\end{equation}
for which we have $\zeta=7/15$. We recall that the smoothness of the path influences the asymptotic scaling of adiabatic state preparation.

\begin{figure*}
\begin{center}
\includegraphics[scale=0.23]{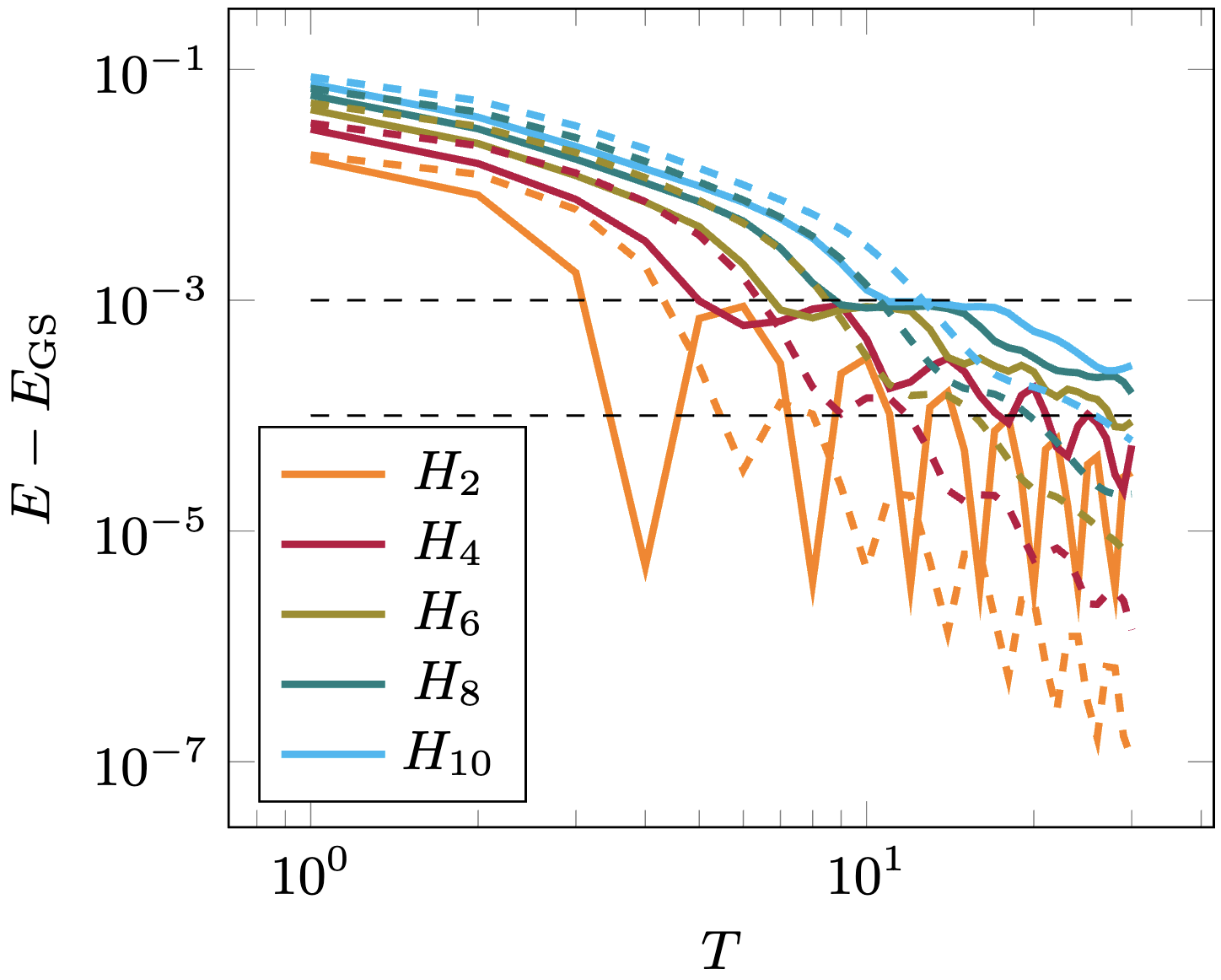}
$\qquad\qquad$
\includegraphics[scale=0.23]{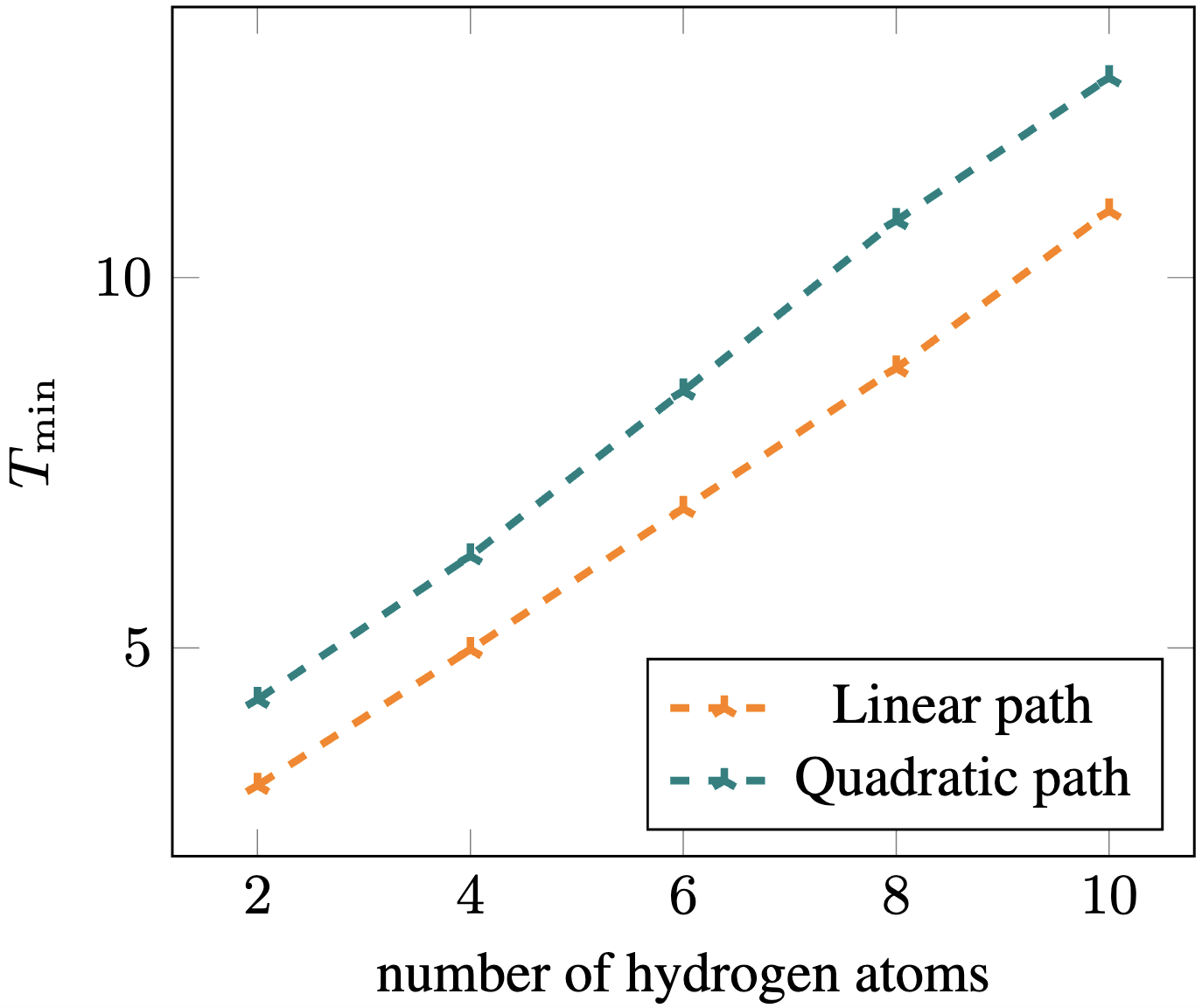}

\caption{\emph{Left panel:} Difference between the ground state energy $E_{\rm GS}$ and the energy $E$ of the state adiabatically evolved up to time $T$, as a function of $T$, for hydrogen chains $H_2,H_4,H_6,H_8,H_{10}$, and two different paths: $w(u)=u$ (continuous lines) and $w(u)=2u^2-u^4$ (dashed lines). The gray dashed line indicate precisions $10^{-3}$ and $10^{-4}$. \emph{Right panel:} lowest adiabatic times $T_{\rm min}$ for which precision $10^{-3}$ is reached in a hydrogen chain, as a function of the number of hydrogens, for linear and quadratic paths.
}
\label {path}
\end{center}
\end {figure*}

In the left panel of Figure \ref{path}, we first compare the energy $E(T)$ reached after an adiabatic evolution during a time $T$, with the ground state energy $E_{\rm GS}$ obtained with exact diagonalization, for a hydrogen chain with $2$ to $10$ atoms in a STO-3G basis. These plots implement the exact adiabatic evolution, without any Trotter error or discretization error. We observe a general decrease of $E(T)-E_{\rm GS}$ as $\sim 1/T^2$ for the linear path, and around $\sim 1/T^4$ for the quadratic path. This is consistent with the fact that smoother paths are associated with better asymptotic scaling. From these curves, we plot in the right panel of Figure \ref{path} the smallest adiabatic times $T_{\rm min}$ for which chemical precision $10^{-3}$ is reached. For both the linear and quadratic paths this time is seen to vary \emph{linearly} with the number of hydrogen atoms in the chain.

\begin{figure*}
\begin{center}
\includegraphics[scale=0.23]{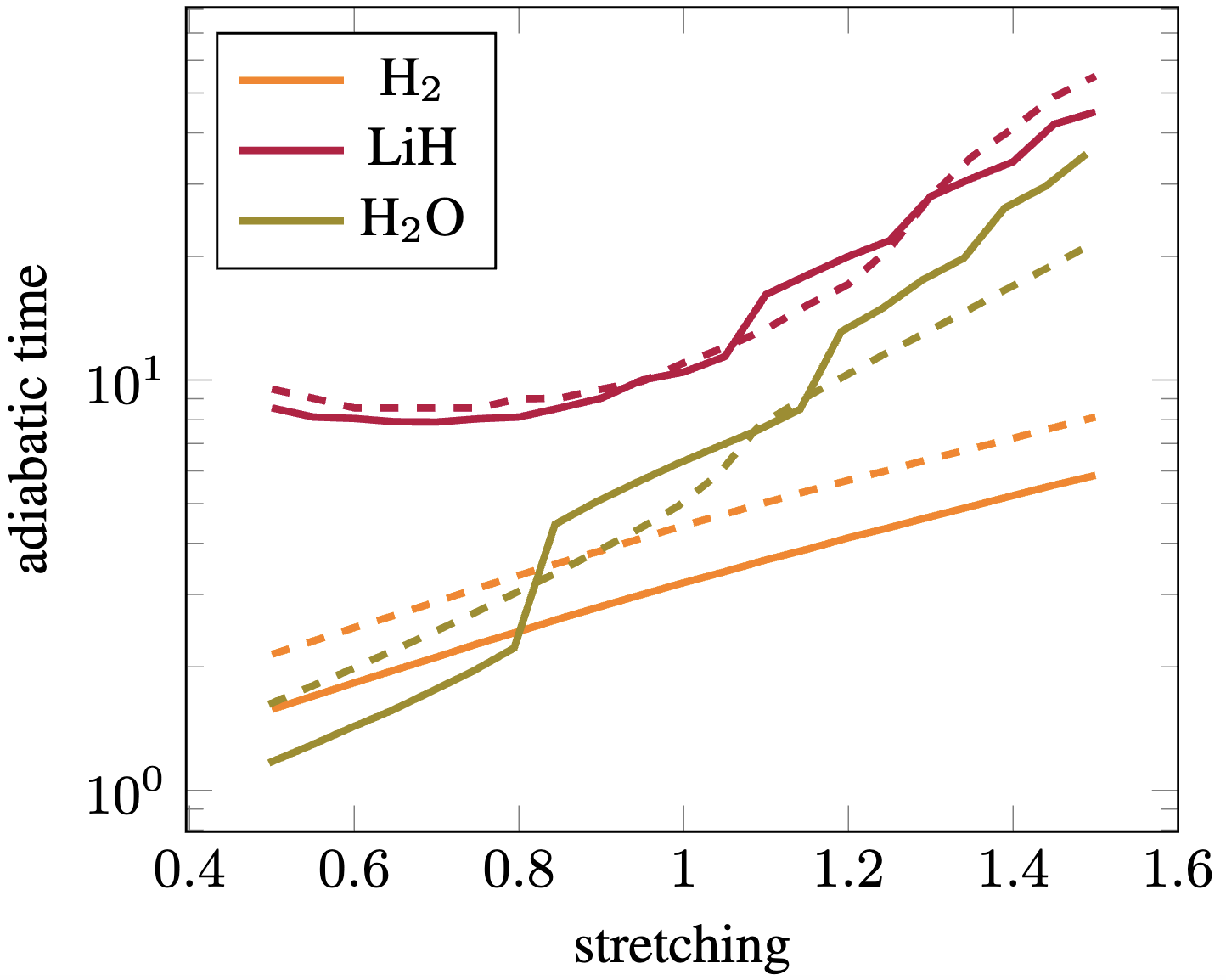}
$\qquad\qquad$
\includegraphics[scale=0.23]{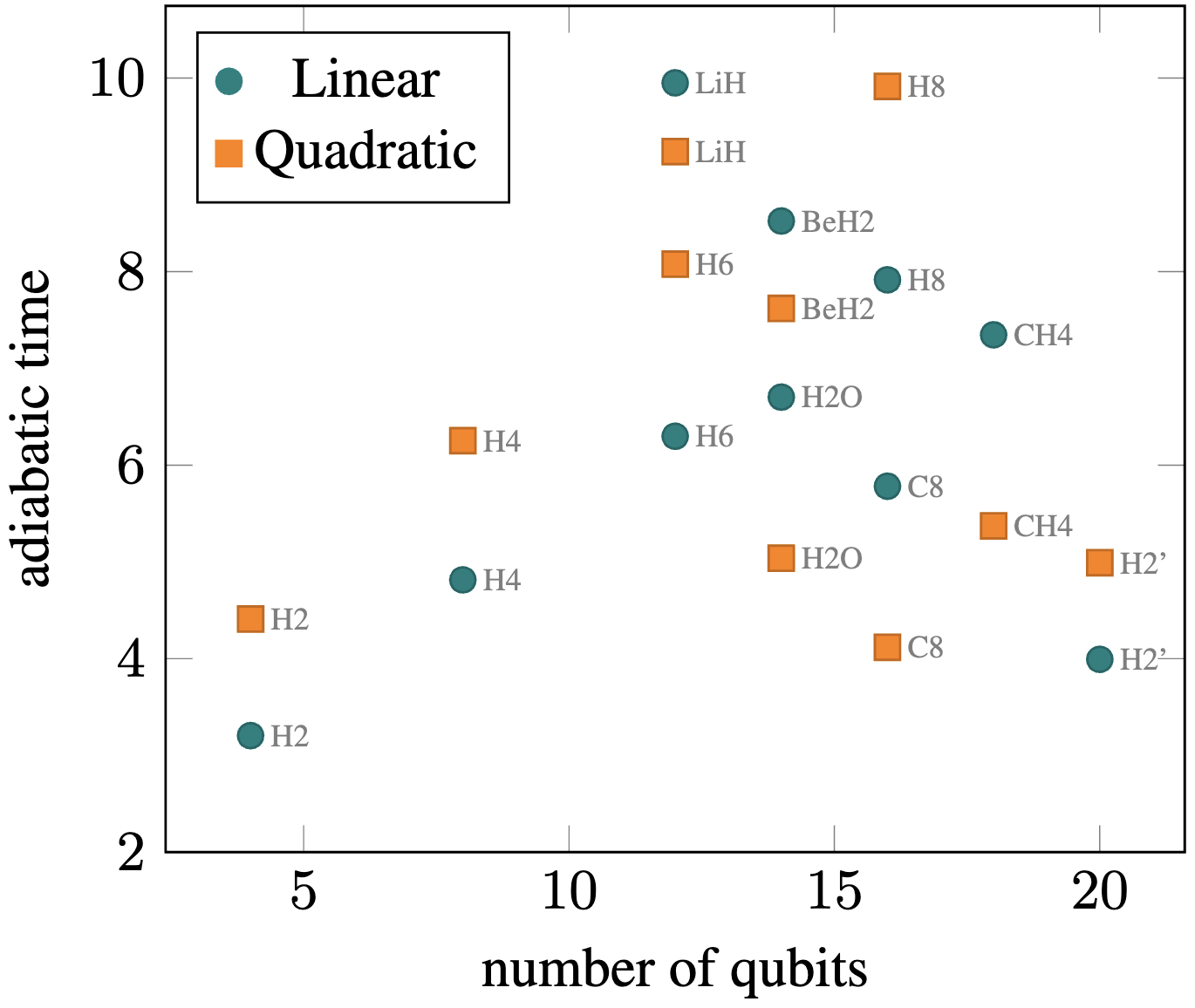}
\caption{\emph{Left panel:} Minimal adiabatic time $T_{\rm min}$ required to reach precision $10^{-3}$ on the ground state energy, using the linear path \eqref{linearweight} (teal, circles) and quadratic path \eqref{quadraweight} (orange, squares), as a function of the number of qubits, for a subset of molecules appearing in Figure \ref{reducednorms}. \emph{Right panel:} Same as left panel, but as a function of the stretching defined as the ratio of the bond length with the equilibrium value. These adiabatic times for stretched geometries are obtained with the standard paths we consider in this paper, but they can be significantly decreased by modifying the path, see Section \ref{pathmodif}.}
\label {adiabatictimes}
\end{center}
\end {figure*}

In the left panel of Figure \ref{adiabatictimes}, we then consider a larger class of molecules for which we compute the minimal adiabatic time required to reach accuracy $10^{-3}$ on the ground state, and plot it as a function of the number of qubits. For most of these molecules that are reachable with exact diagonalization ($\lessapprox 20$ qubits), we observe that the adiabatic time remains below $10$, and can be bounded by $L$ the number of qubits. This is in contrast with adiabatic times that were hypothesised to blow up exponentially with the number of qubits in certain previous works \cite{lee2023evaluating}. On the other hand the order of magnitude of most of the adiabatic times we found is in agreement with certain previous estimates \cite{kremenetski2021simulation}.
In the right panel of Figure \ref{adiabatictimes}, we then consider non-equilibrium geometries by stretching the bond length of some of these molecules. For small stretching we observe only a mild variation of the minimal adiabatic times. However, we see that for larger stretching there can be a significant increase of the adiabatic time with the particular path we study, starting from only single $Z$ Pauli strings.

\subsubsection{2. Path modification}
\setlabel{V.B.2}{pathmodif}
Besides the cases where the adiabatic times needed to reach chemical precision remain small, we also encountered a few cases not listed in Figure \ref{adiabatictimes} where these times become very large. For example, equilibrium hydrogen H${}_2$ in a CCPVDZ basis set with $20$ qubits in our setup requires an adiabatic time $>1000$ to reach chemical precision. These times are obviously very large and could preclude using ASP to solve chemical systems. We also saw in the right panel of Figure \ref{adiabatictimes} that adiabatic times tend to grow fast with the stretching.

One advantage of ASP is the very large freedom one has in the choice of the adiabatic path. Besides varying the speed along the path, one can also vary the initial Hamiltonian and include so-called counter-diabatic terms \cite{demirplak2003adiabatic,berry2009transitionless,chen2010shortcut,hayasaka2023general,hegade2022digitized,sels2017minimizing,claeys2019floquet,jarzynski2013generating}. ASP can be slow for one such path and considerably faster for another path. Let us take the example of equilibrium H${}_2$ in a CCPVDZ basis set, which is $20$ qubits. As said above, when starting from the Hamiltonian containing only single Pauli $Z$ matrices, the adiabatic time required to reach chemical precision is larger than $1000$. If now we take the initial Hamiltonian $H_B$ as the one containing all Pauli strings in $H$ with either exactly one or two $Z$ Pauli matrices (instead of just those with one $Z$), then the adiabatic time to reach chemical precision is reduced to around $4$. A very simple modification of the path can thus completely change the adiabatic time required.

\begin{figure}
\begin{center}
\includegraphics[scale=0.28]{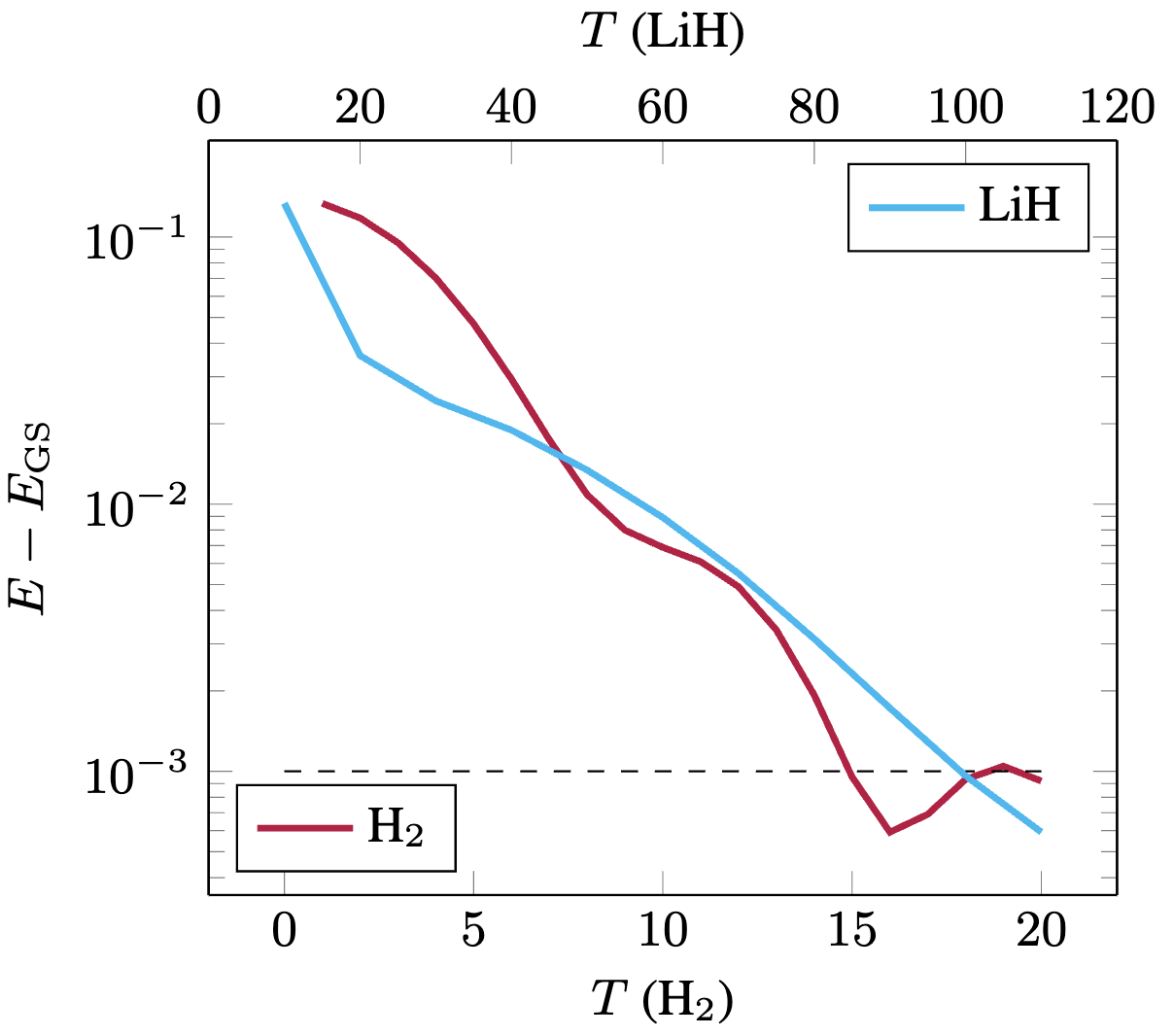}
\caption{Difference between the ground state energy $E_{\rm GS}$ at stretching value $3$ and the energy $E$ of the state adiabatically prepared evolved up to time $T$, as a function of $T$, for H${}_2$ and LiH in a STO-3G basis. The adiabatic path starts from the ground state with stretching $1$ (i.e. equilibrium) and increases the stretching up to $3$, during a total time $T$. The gray dashed line indicates chemical precision $10^{-3}$.}
\label {adiastretch}
\end{center}
\end {figure}

Let us give another example with the non-equilibrium geometries at large stretching, for which ASP was observed to be slow when starting from the Hamiltonian with only single $Z$ terms. We propose instead the following adiabatic path. We first prepare the ground state of the equilibrium geometry by starting from the Hamiltonian with only the single $Z$ terms. As we saw, the minimal adiabatic times remain small for most of the molecules in that setup. Then, we slowly increase the stretching with time $s(t)=s_{\rm initial}(1-\tfrac{t}{T})+s_{\rm final}\tfrac{t}{T}$, performing the Hartree-Fock optimization of the Hamiltonian with that new value of stretching at each time step. We report in Figure \ref{adiastretch} the energy precision obtained during this adiabatic process when preparing the ground state of a molecule with stretching $3$, starting from the ground state of the equilibrium geometry (i.e. stretching $1$), for H${}_2$ and LiH. We observe that this path yields much lower adiabatic times to reach chemical precision. Compared to the previous path, for H${}_2$, this reduces the adiabatic time from $\approx 300$ to $\approx 15$, while for LiH the reduction is from $\approx 800$ to $\approx 100$. 

We note that there exist still several other ways of reducing the adiabatic times or the circuit depth \cite{veis2014adiabatic,keever2023towards}. These examples show that the flexibility of ASP (in the choice of the path, the initial Hamiltonian, possible counter-diabatic terms) can overcome unfavorable setups where adiabatic times needed naively appear to be very large.

\subsection{C. Sampling costs}
\setlabel{V.C}{samplingcosts}
The algorithm presented in Section \ref{adiabatic} is a randomized algorithm, in the sense that it requires to average over several different random circuits to obtain the desired quantum expectation value within the adiabatically evolved state. Compared to a deterministic algorithm like a Trotter decomposition where only one circuit is required, this seemingly comes with a sampling cost overhead. 

If these random circuits were simulated classically, there would be indeed an overhead in including multiple samples to compute the statistical average over the random circuits. However, on a quantum computer, expectation values are computed through averaging measurement outcomes over several realizations of the same circuit, called \emph{shots}. To obtain a precision $\epsilon$ on the expectation value, one has to run in any case of order $\epsilon^{-2}$ shots. In our case, because of the randomized algorithm we use, imprecision on the expectation values comes from two sources: the finite number of random circuits, and the finite number of shots per circuit. Given a total shot budget $S$, it is a priori non trivial how to best distribute these $S$ shots among different random circuits, i.e., what is the optimal number of shots $s$ per circuits that minimizes the total variance over the $S/s$ random circuits. As shown in \cite{granet2023continuous}, this optimal number is $s=1$, namely one single shot per circuit. In that case the noise coming from the finite number of circuits and that coming from the finite number of shots per circuit are indistinguishable. 

Let us now estimate the actual statistical variance associated with the random circuit drawing in the case where we do only one shot per circuit. We write an amplitude as
\begin{equation}
    \langle 0|\mathcal{A}(T)^\dagger e^{is(E-H)} \mathcal{A}(T)|0\rangle=(\lambda_A^2 \lambda_C)^{-1}\mathbb{E}[\langle 0| U_1^\dagger U_1' U_2 |0\rangle]\,,
\end{equation}
with $U_1,U_2$ two independent random circuits drawn to generate $\mathcal{A}(T)$, and $U'_1$ a random circuit drawn to generate $e^{is(E-H)}$. $\lambda_A$ and $\lambda_C$ are the attenuation factors corresponding to generating $\mathcal{A}(T)$ and $e^{is(E-H)}$ respectively. We will denote $\lambda=\lambda_A^2\lambda_C$ the total attenuation factor. For $-1\leq x\leq 1$, we denote $\mathcal{B}(x)$ the Bernoulli random variable that takes value $+1$ with probability $\tfrac{1+x}{2}$ and $-1$ with probability $\tfrac{1-x}{2}$. We denote then $X$ the random variable
\begin{equation}
    X=\lambda^{-1}\mathcal{B}(\Im \langle 0| U_1^\dagger U_1' U_2 |0\rangle)\,.
\end{equation}
This random variable is exactly the measurement outcome of one shot of a circuit that measures the imaginary part of $\langle 0| U_1^\dagger U_1' U_2 |0\rangle$, multiplied by $\lambda^{-1}$. We thus have the mean value $\mathbb{E}[X]=\Im  \langle 0|\mathcal{A}(T)^\dagger e^{is(E-H)} \mathcal{A}(T)|0\rangle$. The number of circuits to include to converge to the mean is around the variance $\approx {\rm Var}(X)$ with
\begin{equation}
    {\rm Var}(X)=\mathbb{E}[X^2]-\mathbb{E}[X]^2\,.
\end{equation}
In our case, since $\mathcal{B}(x)^2=1$, this variance is exactly
\begin{equation}
    {\rm Var}(X)=\frac{1}{\lambda^2}-(\Im  \langle 0|\mathcal{A}(T)^\dagger e^{is(E-H)} \mathcal{A}(T)|0\rangle)^2\,.
\end{equation}
Hence the statistical variance of the algorithm is bounded by the inverse square of the total attenuation factor
\begin{equation}
     {\rm Var}(X)\leq \frac{1}{\lambda^2}\,.
\end{equation}
This means we need at most $\approx 1/\lambda^2$ shots to converge to precision $\mathcal{O}(1)$, and $\approx 1/(\lambda\epsilon)^2$ to converge to precision $\epsilon$. Thus, in contrast with Trotterization (where the actual amplitude of Trotter errors are a priori unknown), for any desired confidence interval, the required total cost (i.e. number of gates per sample times number of samples) can be known in advance.

\begin{figure}
\begin{center}
\includegraphics[scale=0.23]{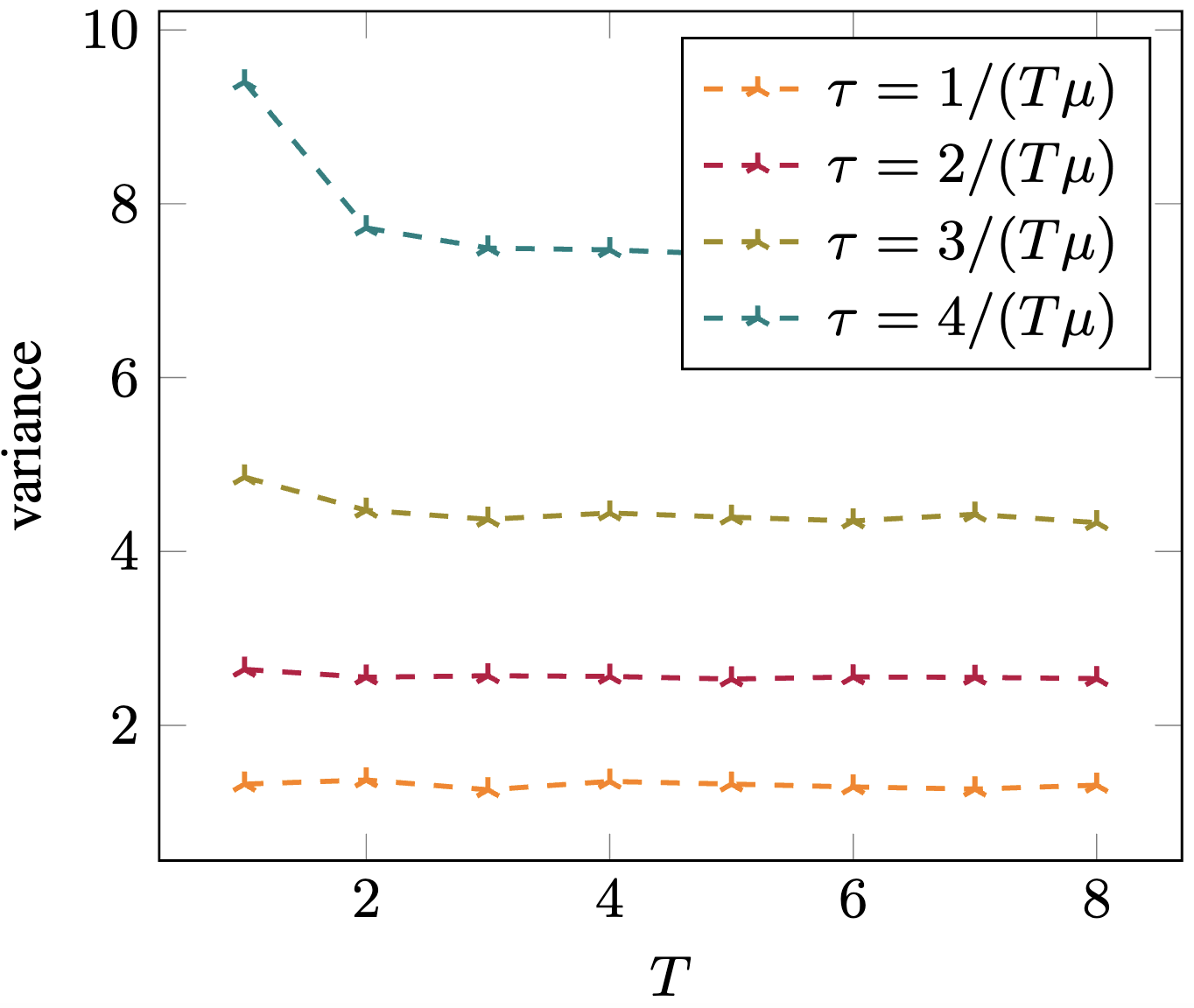}
\caption{Statistical variance in measuring $\langle 0|\mathcal{A}(T)^\dagger \mathcal{A}(T)|0\rangle=1$ with the randomized algorithm with $1$ shot per circuit, as a function of $T$, for different gate angles $\tau(T)$, for a H${}_4$ molecule at equilibrium in a STO-3G basis.}
\label {samplingcostsfigure}
\end{center}
\end {figure}

In Figure \ref{samplingcostsfigure}, we evaluate numerically the variance in the case $s=0$, measuring the real part of the amplitude (which should be $1$) instead of the imaginary part. We fix the gate angle $\tau=\frac{n}{T\mu_I}$ for $n=1,2,3,4$ and vary $T$. For this value of the gate angle the attenuation factor should become independent of $T$ as $T$ grows, with value $ \lambda^{-1} \sim e^{n}$. We observe indeed this behaviour. This shows that the randomized nature of the algorithm, provided the optimal gate angle $\tau\sim \frac{1}{T\mu_I}$ is chosen, does not bring any overhead compared to a deterministic algorithm like Trotter.

\subsection{D. Discretization costs of Trotter}
\setlabel{V.D}{Discretization}

\begin{figure*}
\begin{center}
\includegraphics[scale=0.23]{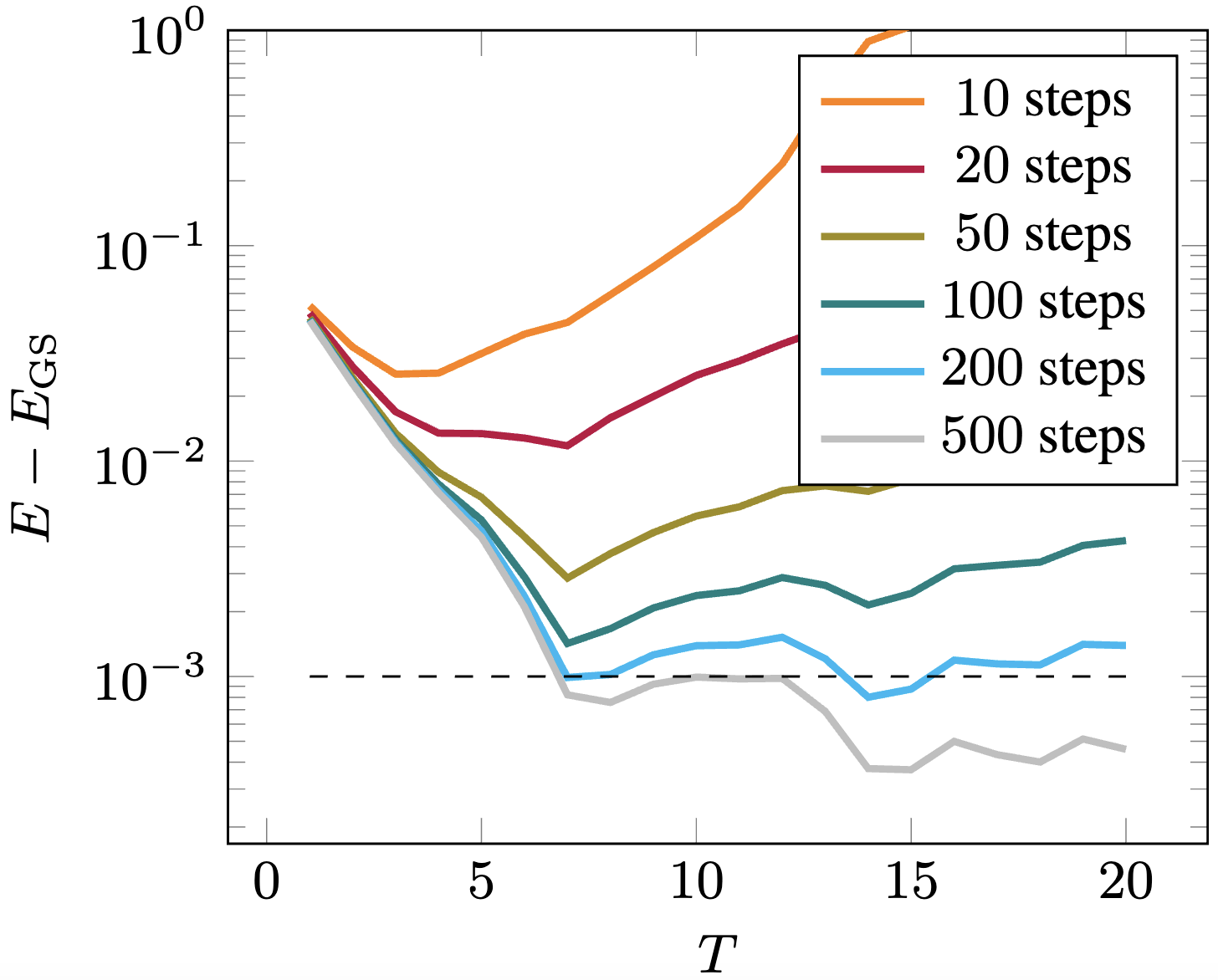}
$\qquad \qquad$
\includegraphics[scale=0.23]{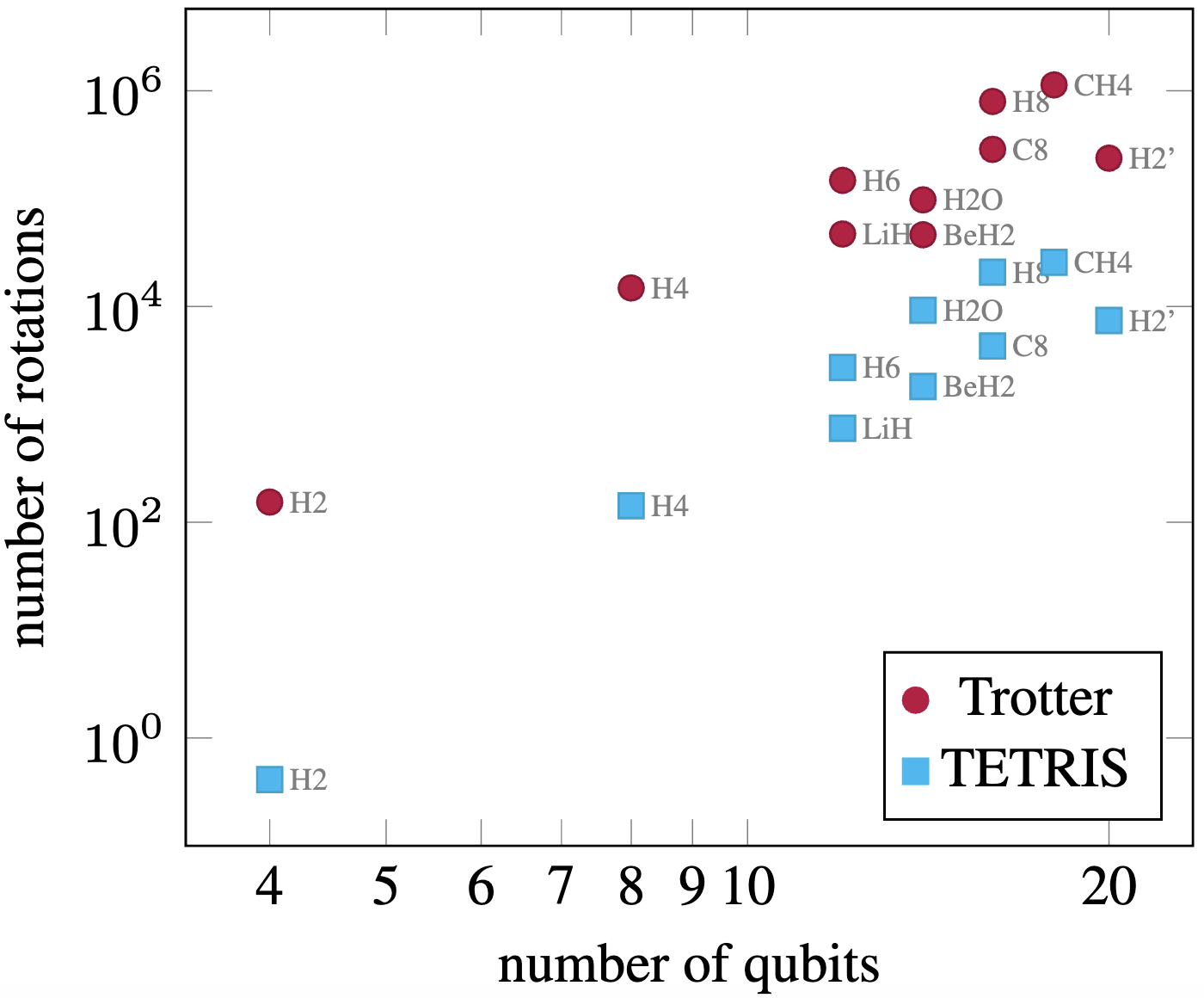}
\caption{\emph{Left panel:} difference between energy of the state prepared and ground state energy, as a function of adiabatic time $T$ for different number of Trotter steps with linear adiabatic path, for the H${}_6$ hydrogen chain. \emph{Right panel:} Minimal number of Pauli strings rotations $e^{i\tau P_n}$ (of length $\geq 2$) to perform to reach chemical precision by adiabatically preparing the ground state with a linear path, using a Trotter implementation (purple, circles) and a TETRIS implementation $\frac{1}{2}T^2\mu_I^2$ (cyan, squares), for different molecules.}
\label {adiabatictimes2}
\end{center}
\end {figure*}

Let us now evaluate the cost of implementing ASP with a Trotter decomposition, and compare with our proposal. To use the Trotter algorithm to implement ASP, we have to approximate the adiabatic evolution operator as
\begin{equation}\label{approxtrotter}
    \mathcal{A}(T)\approx e^{i\Delta t H(0)}e^{i\Delta t H(\Delta t)}...e^{i\Delta t H(T-\Delta t)}\,,
\end{equation}
with $\Delta t$ a chosen time step, and $H(t)$ the time-dependent Hamiltonian of interest, evaluated at time points $0,\Delta t,2\Delta t,...$. Then each of the term $e^{i\Delta t H(u)}$ at time $u$ is itself written as a product of exponentials $e^{i\Delta t H(u)}\approx \prod_{n=1}^Ne^{i\Delta t c_n(u)P_n}$, with $c_n(u)$ the corresponding coefficient of $H(u)$ in front of Pauli string $P_n$. This implementation comes both with a discretization error (the first approximation) and a Trotter error (the second approximation), that both lead to a heating, namely to an increase of the energy of the state prepared. In the left panel of Figure \ref{adiabatictimes2}, we plot the energy of the adiabatic evolution as a function of the final time $T$, for different fixed number of steps $1/\Delta t$, for a H${}_6$ hydrogen chain. The heating effect is clearly visible on the plot. At a fixed finite number of Trotter steps, when increasing the total simulation time $T$, there is inevitably an increase of the energy of the state prepared after some time $T$. This precludes reaching chemical precision if the number of Trotter steps is too low. In this particular example we see that we require $\sim 200$ time steps to reach chemical precision.
Since each Trotter step requires to implement a rotation $e^{i\tau P_n}$ for the $919$ terms of that particular molecule, this yields $\sim 10^5$ rotations. 

When implementing a Trotter decomposition, one has to choose the order of the terms, and different orderings may display more or less pronounced discretization error \cite{childs2019faster}. However, for these large number of Trotter steps required to reach chemical precision, these variations are observed to be negligible. For example, for H${}_6$ at time $T=7$ with $300$ Trotter steps, by permuting randomly the terms in the Trotter decomposition we find energies in the range $0.98\pm 0.03$mH above the ground state. The curves in the left panel of Figure \ref{adiabatictimes2} would thus be similar for different Trotter orderings.

In contrast, in our case with the algorithm of Section \ref{adiabatic}, we \emph{do not} have any discretization error. In particular, we do not have the approximate decomposition of \eqref{approxtrotter}: ASP is implemented exactly. For the example of the H${}_6$ hydrogen chain mentioned above, the algorithm we use requires $\frac{1}{2}\mu^2_I T^2$ rotations where $\mu_I=11.7$ and $T=7$, which is around a factor $10^2$ less than a Trotter implementation. Moreover, by multiplying the gate angle $\tau$ (which is a free parameter of the algorithm) by some factor $\kappa>0$, one can always divide the depth of the circuit by $\kappa$, provided we can afford a shot overhead $\sim e^{\kappa}$. In the right panel of Figure \ref{adiabatictimes2}, we compare the number of rotations $e^{i\tau P_n}$ that have to be implemented to reach chemical accuracy using a Trotter decomposition and with TETRIS, for several different molecules with a number of orbitals within reach of exact diagonalization. We observe a significant reduction of order $10^2$ with TETRIS in the number of gates per circuit. This shows that the algorithm of Section \ref{adiabatic} is able to significantly reduce the cost of ASP.

\subsection{E. Cost of other time evolution algorithms}
As said in the introduction, there exist many other techniques to implement Hamiltonian evolution besides Trotter. Let us briefly comment on them. Another randomized algorithm for Hamiltonian time evolution is the qDRIFT algorithm \cite{campbell2019random}. The algorithm has a gate count $\mathcal{O}(t^2\mu^2/\epsilon)$ for time $t$, $1$-norm $\mu$ and precision $\epsilon$. However, it is known to have a large prefactor and to present important discretization error, see e.g. \cite{granet2023continuous}. There exist as well time evolution algorithms with more favorable scaling than Trotter, for example Linear Combination of Unitaries (LCU) \cite{berry2015simulating}. This algorithm has a gate count $\mathcal{O}(N\mu t \tfrac{\log \mu t/\epsilon}{\log \log \mu t\epsilon})$, with $N$ the number of terms in the Hamiltonian. Approximating the prefactor and the log term by some constant $C$, the cost is thus $C L^4\mu T$. This prefactor is known to be large. It follows that TETRIS has a better performance when
\begin{equation}
   T<C\frac{L^4}{\mu}\,.
\end{equation}
Using the scaling of $\mu$ that we obtained, this is approximately
\begin{equation}
    T\lessapprox 100 C L\,.
\end{equation}
This condition is largely satisfied for the molecules we studied in Figure \ref{adiabatictimes}. We note that by changing the orbital basis to minimize the $1$-norm of the Hamiltonian, this condition could be brought to $T\lessapprox \mathcal{O}(L^2)$ \cite{koridon2021orbital}. This shows the better practicality of TETRIS for quantum chemistry applications at least in the near term.

\subsection{F. Scaling analysis}
To conclude this Section, we perform a basic scaling analysis of the cost of ASP for chemistry applications, based on the previous resource estimates. Using $g=L/2$, $\zeta=1/2$, adiabatically preparing a state $|\psi\rangle \langle \psi|$ has a cost 
\begin{equation}
    C_{\rm prepare}=L\mu_I^2T^2\,.
\end{equation}
The cost of measuring the energy with precision $10^{-3}$ using the different methods presented in Section \ref{measuringenergy} is
\begin{equation}
    C_{\rm measure}=10^6 L \mu^2_I\,.
\end{equation}
For the hydrogen chain, we had the estimates $\mu_I\approx 0.2 L^{2.13}$ and adiabatic times $T\approx L/2$. This yields the scaling
\begin{equation}\label{scalingadia}
    C_{\rm total}\sim 10^{-3} \cdot L^{6.6}(10^3+L)^2\,.
\end{equation}
This estimate shows that for systems with less than $\sim 10^3$ qubits, the measurement of the energy is the most costly step, whereas for larger systems the adiabatic preparation step is more costly.

For more generic molecules, we had  the estimate $\mu_I\approx 0.007 L^{2.8}$. The adiabatic times required are more difficult to estimate, but on the molecules we tested at equilibrium we found an upper bound $T\approx L$. Hence in total, we get a rough upper bound estimate of the total number of CNOTs
\begin{equation}\label{scalingadia2}
    C_{\rm total}\sim 10^{-3} \cdot L^{6.6}(10^3+L)^2\,.
\end{equation}
We again recall that these estimates are obviously very dependent on the scaling of the adiabatic times $T$ required to reach chemical precision, that we estimated for systems with up to $20$ spin orbitals.

\section{VI. Noisy simulations}
\setlabel{VI}{simulationssections}

\subsection{A. Setup}

We now present numerical simulations of the algorithm that take into account hardware imperfections. We compile every circuit into Pauli matrices, $S,S^\dagger,H$, $Z$ rotations, and the two-qubit gate $e^{i\theta Z_1Z_2}$, which is the native coupling gate in Quantinuum's ion-trap hardware \cite{moses2023racetrack}. We model hardware noise by a simple depolarizing channel after every $e^{i\theta Z_1Z_2}$ gate.
This kind of basic noise model gives a good approximation of many current platforms.

\subsection{B. Symmetry filtering}
\setlabel{VI.B}{symmetryfilter}
The presence of symmetries in a system can be exploited to reduce the noise on the measurement outcomes of hardware or noisy simulations. If the unitary operator that a circuit implements commutes with some charge operator $Q$, then one can discard in the measurement outcomes all the shots where the value of $Q$ is modified compared to the initial value of $Q$. 

In our case, the Hamiltonians describing molecules always conserve the number of electrons with spin up or down separately. However, the use of the randomization algorithm comes with a subtlety. Although the total Hamiltonians conserve particle number $Q=\sum_{i=1}^N Z_i$, each term in their Pauli string decomposition \emph{does not} necessarily conserve particle number. After the Jordan-Wigner transformation, we obtain terms like $X_1X_2+Y_1Y_2$ which commutes with particle number $Z_1+Z_2$. However, in every random circuit generated by TETRIS, rotations with respect to $X_1X_2$ and $Y_1Y_2$ are performed separately, and these terms separately do not commute with $Z_1+Z_2$ and do not conserve particle number. When drawing randomly the gates, the circuits that we obtain thus \emph{do not individually} conserve particle number, although \emph{on average} they will. Even in the noiseless case, some shots will not conserve particle number. Discarding them would lead to a biased result. However, since we know these shots have to average to $0$ in the result, one can impose that these non-particle-conserving shots contribute as zero in the average over the shots (which is different from discarding them, because the total number of shots that we use to average the results remains the same). In the presence of noise, shots that do not conserve particle number can come from either the fact that the circuit itself does not conserve particle number, or from the fact that an error has occurred. However, because non-particle-conserving shots can also occur in noiseless circuits, we cannot discard these shots in the noisy case, and one can only impose that they contribute as zero in the result. The consequence is that this symmetry filtering \emph{does not} remove noise, but only converts it into a global depolarizing channel that maps the final density matrix $\rho$ into $(1-\lambda)\rho+\frac{\lambda}{2^L}$ with some noise level $\lambda$.

There is however a less constraining symmetry that is satisfied at the level of each random circuit generated by the algorithm. This is the \emph{parity} of particles with spin up or down. As this property is conserved individually in each circuit, no shots breaking the symmetry will be observed in the noiseless case, and one can systematically discard non-parity-conserving shots in hardware data or noisy simulations as one is certain that an error has occurred in those shots. This is the technique that we will use below when any symmetry filtering is mentioned.

\subsection{C. Noise-resilience}
\setlabel{VI.C}{noiseresinum}
In Section \ref{noiseresi} we proposed three different methods to extract the ground state energy from the measurement of the expectation value of the operator $e^{is(E-H)}$ within the state prepared adiabatically. If the state prepared is sufficiently close to the ground state, this amplitude  is $\sin(s(E-E_{\rm GS}))$ with $E_{\rm GS}$ the desired ground state energy. We argued that these methods are noise-resilient because the main effect of hardware noise is to dampen the signal $\sin(s(E-E_{\rm GS}))$ by multiplying it by a positive constant that is independent of $E$.

\begin{figure}
\begin{center}
\includegraphics[scale=0.23]{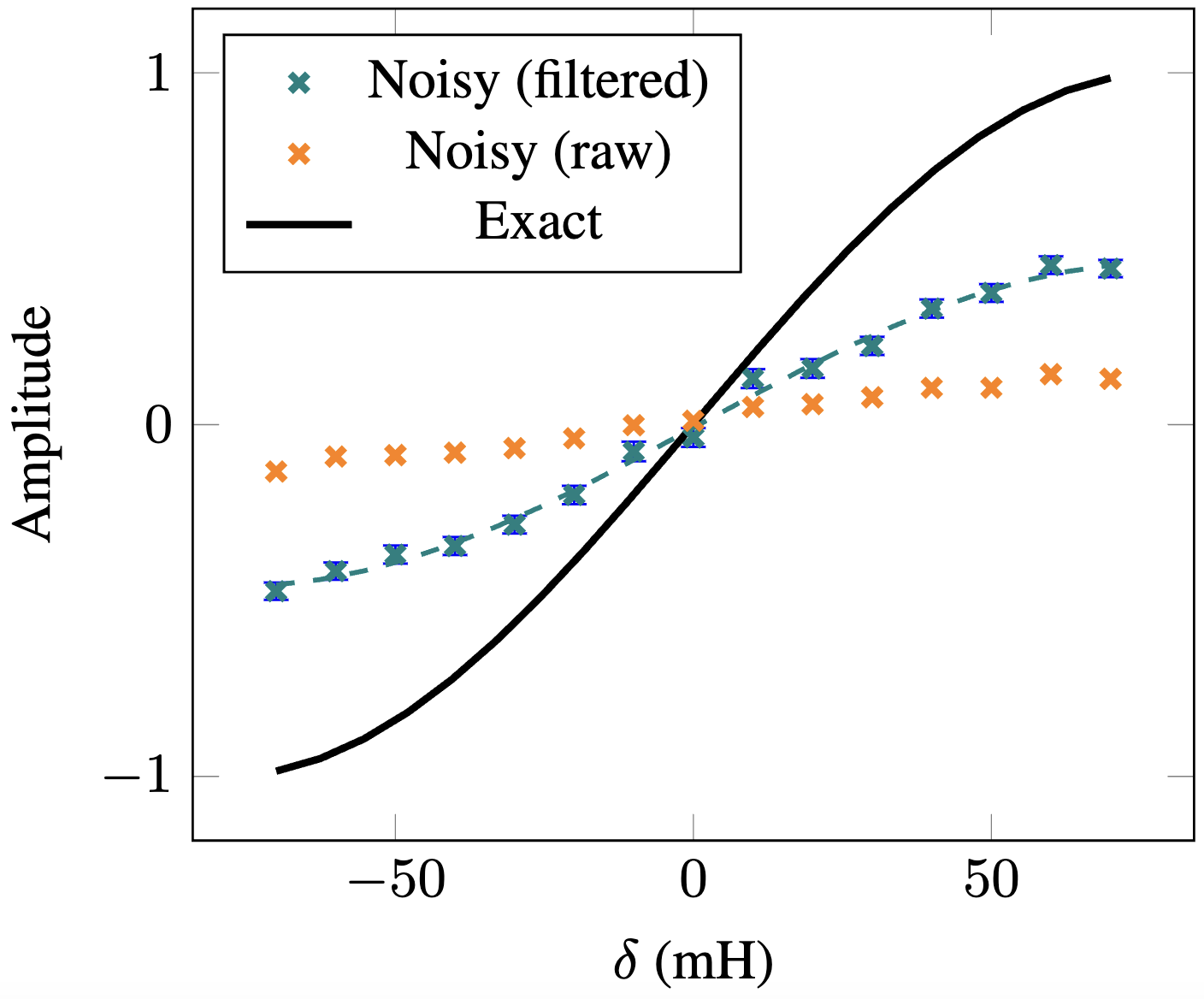}
\caption{Imaginary part of the expectation value of $e^{is(E-H)}$ in the state prepared with ASP, for a H${}_2$ molecule with bond length $1.11$\r{A}, using an adiabatic time $T=12$ with a linear path, as a function of $\delta=E-E_{\rm GS}$ expressed in mH, at fixed $s=20$: exact curve $\sin(0.02\delta)$ (continuous black), raw noisy simulations (orange) and noisy simulations after parity filtering (teal). The noise is modelled by a depolarizing channel with amplitude $0.01$ after every two-qubit gate. The dashed teal curve is a fit $0.46 \sin(0.02(\delta-0.979))$.}
\label {noise0}
\end{center}
\end {figure}

In this Section, we start by checking this affirmation with a very common noise model where a depolarizing channel on two qubits is applied after every two-qubit gate. In Figure \ref{noise0} we show simulations for a stretched hydrogen ${\rm H}_2$ molecule in a STO-3G basis with bond length $1.11$\r{A}, on $4$ qubits, comparing the noisy simulations of $\Im\langle \psi|e^{is(E-H)}|\psi\rangle$ with exact value, using depolarizing channels with amplitude $0.01$ and the corresponding optimal gate angle $\tau$ as described in Section \ref{optimalnoise}. We show both the raw data and the data obtained after parity filtering as introduced in Section \ref{symmetryfilter}. For these parameters there are around $200$ two-qubit gates per circuit. We first observe that the raw noisy data are hugely dampened compared to the exact value. At $\delta=E-E_{\rm GS}=\pm 70$mH, the amplitude is damped by a factor $\approx 0.131$, which is very close to the rule of thumb $0.99^{200}\approx 0.134$, namely the fidelity of every two-qubit gate raised to the number of two-qubit gates in the circuit. The parity filtering is seen to have a large effect and multiplies the amplitude of the signal by a factor $\approx 3-4$. 

Next, we observe that the main effect of the noise after parity filtering is to dampen the curve. The value of $E$ where the amplitude $\Im\langle \psi|e^{is(E-H)}|\psi\rangle$ vanishes (which is $E=E_{\rm GS}$) is barely affected by the noise. By fitting the noisy data points after parity filtering, we find that the results fit well with a sinus. The amplitude is dampened by a factor $0.46$ compared to exact, and the point where the curve vanishes is shifted by only $0.98$mH. This confirms the noise-resilience of extracting the ground state energy from the point where the amplitude $\Im\langle \psi|e^{is(E-H)}|\psi\rangle$ vanishes.

\subsection{D. Performance of the three methods proposed}
\begin{figure*}
\begin{center}
\includegraphics[scale=0.23]{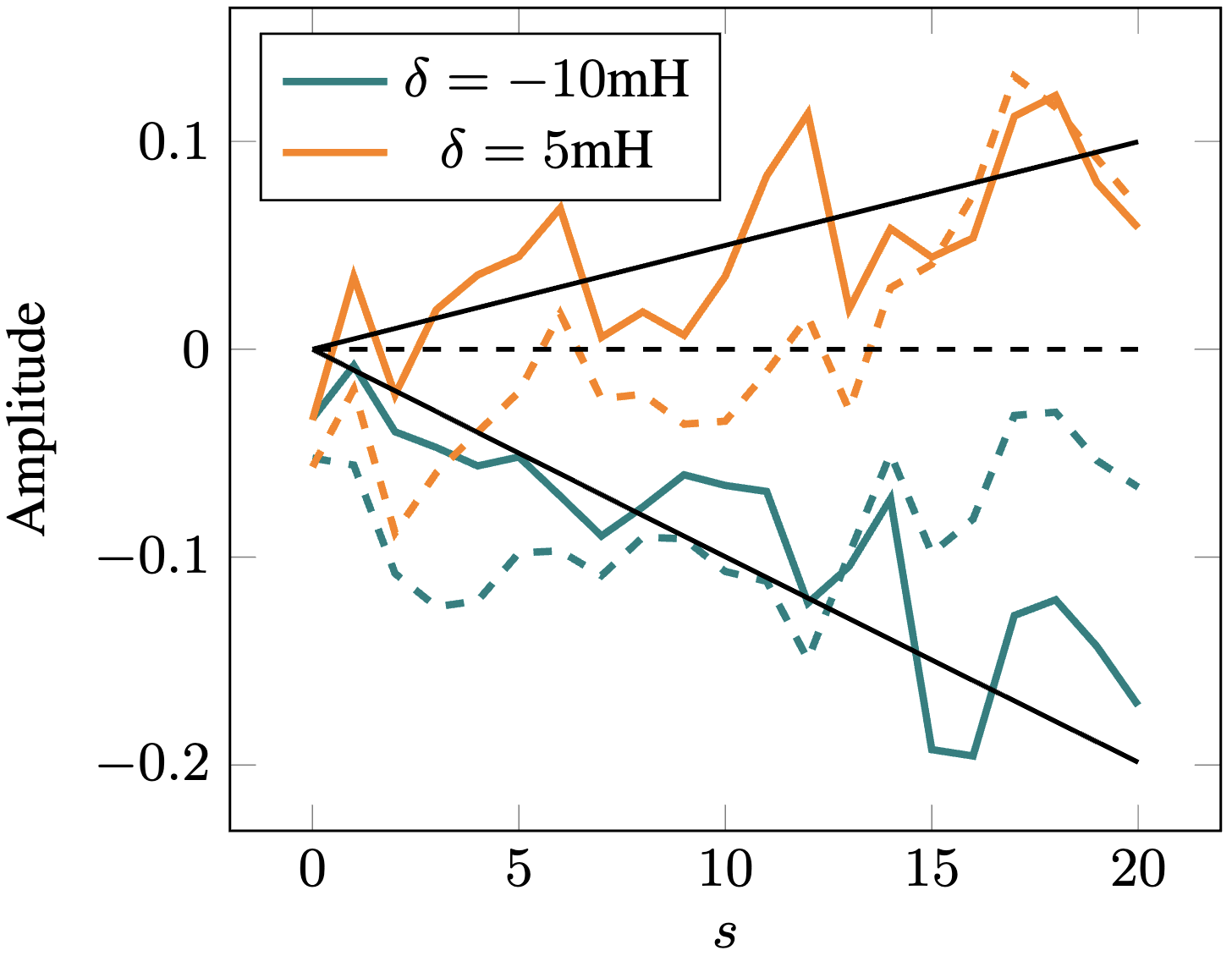}
$\qquad \qquad$
\includegraphics[scale=0.23]{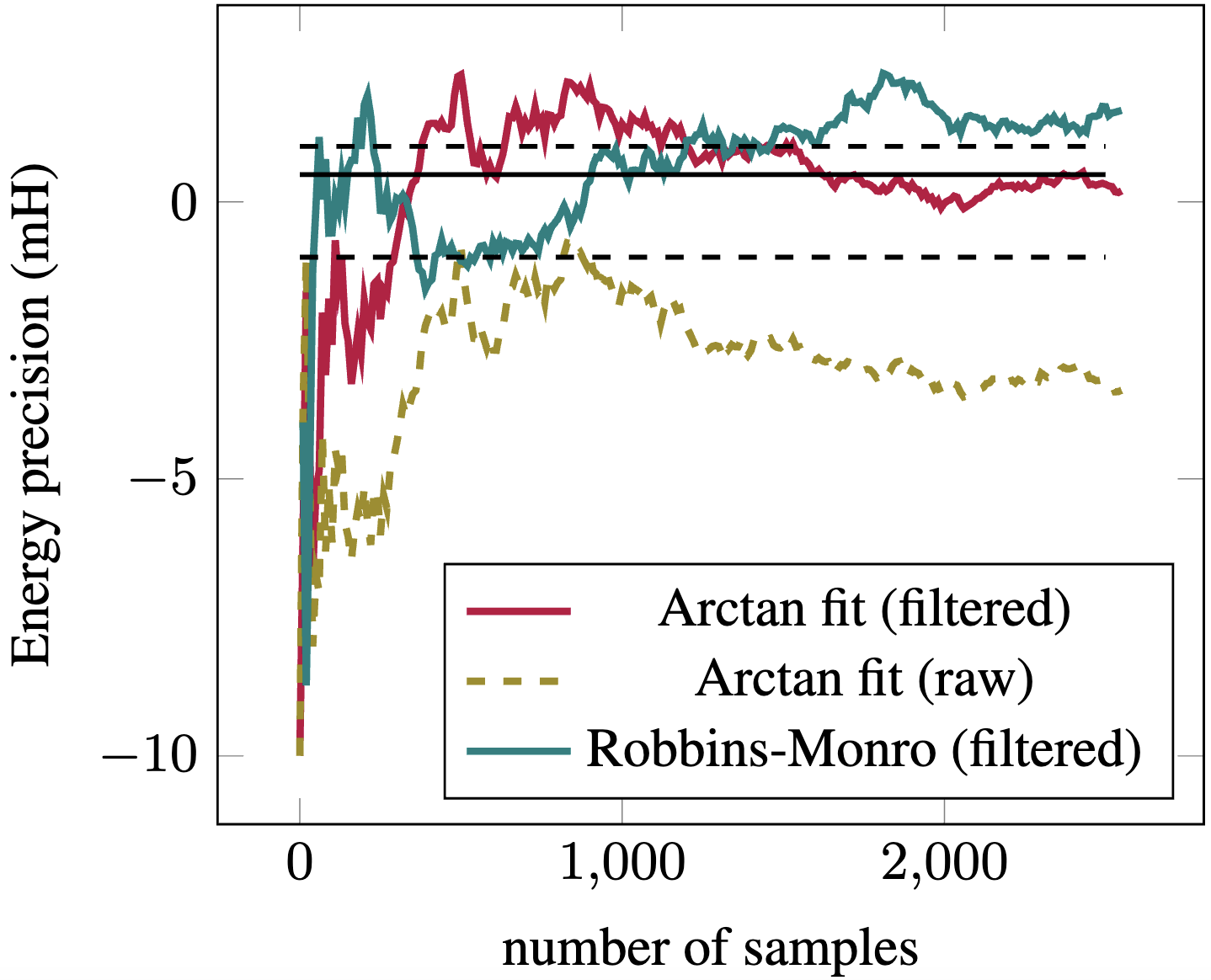}
\caption{\emph{Left panel:} noisy amplitude $\Im \langle \psi(T)|e^{is(E-H)}|\psi(T)\rangle$ as a function of $s$, for a H${}_2$ molecule with bond length $1.11$\r{A}, using an adiabatic time $T=12$ with a linear path, for two different values of $\delta=E-E_{\rm GS}$, for a two-qubit gate fidelity $0.9982$, with $1000$ samples with $100$ shots per sample. Continuous lines indicate values after parity filtering, and dashed lines raw data. The black lines indicate the theoretical noiseless values. \emph{Right panel:} Precision on the estimated energy with the arctan fit formula \eqref{formulalinear} for $s=20$, $E_{test}=E_{\rm GS}-10$mH, and $\epsilon=20$mH, and for the Robbins-Monro approach with $a_n=\frac{10}{n^{0.75}}$ starting at $\delta=-10$mH, as a function of the number of samples considered, in the same conditions as in the left plot. The black continuous line indicates $0.49$mH the energy of the adiabatic state prepared, and dashed lines indicate chemical precision $1$mH.}
\label {noise}
\end{center}
\end {figure*}

In Figure \ref{noise} we show simulations for a stretched hydrogen ${\rm H}_2$ molecule in a STO-3G basis with bond length $1.11$\r{A}, on $4$ qubits. We show the results for the three different approaches to measure the energy presented in Section \ref{measuringenergy}, using the same adiabatic state preparation with linear path and adiabatic time $T=12$. The energy of the state produced this way is $0.49$mH above the exact ground state energy.

In the left panel, we show the results of the binary search approach, where one determines whether the ground state energy is above or below a certain energy $E$ by measuring the sign of $\Im \langle \psi(T)|e^{is(E-H)}|\psi(T)\rangle$. We consider two cases where $E$ is at a distance $-10$mH and $5$mH from the ground state, and plot the measured value of $\Im \langle \psi(T)|e^{is(E-H)}|\psi(T)\rangle$ as a function of $s$ the central time, using $1000$ samples and $100$ shots per sample. We see that the raw unfiltered results are systematically shifted because of noise. This occurs even for $s=0$, because the circuit still contains the adiabatic state preparation in this case. However, after using the parity filter, we observe that up to the shot noise the results are compatible with the noiseless exact values. In particular, they allow one to distinguish positive from negative by going to central times $s\approx 15-20$. Nevertheless, performing the same calculation with a case where $E$ is at a distance $\pm 1$mH from the ground state would be challenging and would require to go to large values of $s$ or much larger number of shots. 

In the right panel, we then test the two other approaches, the arctan fit approach and the Robbins-Monro approach. We start from an initial energy estimate $10$mH below the exact ground state. For the arctan fit approach, we then measure $\Im \langle \psi(T)|e^{is(E-H)}|\psi(T)\rangle$ at energies $30$mH below and $10$mH above the ground state (i.e., $\delta=-10$mH and $\epsilon=20$mH), and plot formula \eqref{formulalinear} as a function of the number of samples. We see that the ground state estimate improves immediately with just one hundred samples, and reaches chemical precision after around $1000$ samples.  As for the Robbins-Monro approach, we use the parameters $a_n=\frac{10}{n^{0.75}}$ and plot the energy iterates \eqref{energyiterates} as a function of the number of iterations. We see again a very fast increase in precision at the beginning. Values close to chemical precision are also reached after around $1000$ samples. Optimal choices of coefficients $a_n$ could improve further the results.

\section{VII. Conclusion}
\setlabel{VII}{conclusion}
In this work we investigated the relevance of adiabatic state preparation for chemistry applications, motivated by the  algorithm in \cite{granet2023continuous} to implement time-dependent Hamiltonian evolution without Trotter heating. 

Firstly (i), we evaluated the cost to adiabatically prepare the ground state of molecules within chemical precision $10^{-3}$, using a Trotter decomposition and with the TETRIS algorithm. We find that TETRIS divides by a factor $\sim 10^2$ the number of two-qubit gates required to reach chemical precision compared to Trotter, for several molecules within reach of exact diagonalization on less than $20$ qubits. Moreover, we evaluated the minimal adiabatic times required to reach chemical precision and found in many cases that they either remain of order $1$ or $10$, or that modifications of the adiabatic path can drastically reduce them. For example, the adiabatic times required for a hydrogen chain scale only linearly with the number of atoms.

Secondly (ii), we proposed three different approaches inspired from QPE to measure the energy of the state adiabatically prepared. Indeed, the precision required in chemistry applications is a technical challenge in terms of noise, number of measurements and runtime. Moreover, because TETRIS requires to average complex amplitudes and not their absolute values squared, one cannot use the efficient and optimized techniques that have been developed to measure efficiently the energy of states prepared with VQE. These methods rely on the fact that when $|\psi\rangle$ is an eigenstate of $H$, the value of $E$ where the imaginary part of $\langle \psi|e^{it(E-H)}|\psi\rangle$ vanishes (that is the eigenvalue of $|\psi\rangle$ on $H$) is resilient to depolarizing noise channels. We tested numerically these approaches on small molecules and showed that they are efficient. 

From these different results we conclude that adiabatic state preparation is a viable approach to compute the ground state energy of molecules within chemical precision. Nevertheless, the absolute costs remain large, and today's or near term quantum computer can only solve small molecules. At the algorithmic level, this suggests the need for further optimization, which could be done in several different directions. We can mention for example modifying the path or the speed along the path, modifying the orbital basis in order to minimize the $1$-norm of the Hamiltonian, starting from a different initial state with lower initial energy. As each of these directions can yield drastic savings, adiabatic state preparation is a serious candidate for solving chemistry problems on a near-fault-tolerant quantum computer.

\section{Acknowledgements}
We thank Ramil Nigmatullin and Kentaro Yamamoto for comments on the draft. E.G. acknowledges support by the Bavarian Ministry of Economic Affairs, Regional Development and Energy (StMWi) under project Bench-QC (DIK0425/01). K.H. and H.D. acknowledge support by the German Federal Ministry of Education and Research (BMBF) through the project
EQUAHUMO (grant number 13N16069) within the funding program quantum technologies - from basic research to market.

%


\end{document}